\documentclass[apj,twocolumn,twocolappendix,numberedappendix]{openjournal}
\usepackage{newtxtext,newtxmath,enumitem,multirow,ctable}
\usepackage[T1]{fontenc}
\usepackage[breaklinks,colorlinks,citecolor=blue,urlcolor=blue]{hyperref}

\newcommand{\gizmourl}{\href{http://www.tapir.caltech.edu/~phopkins/Site/GIZMO.html}{\url{http://www.tapir.caltech.edu/~phopkins/Site/GIZMO.html}}}

\newcommand{\orcidauthor}[3]{\author{\href{http://orcid.org/#1}{#2$^{#3}$}}}

\shorttitle{Time Dilation for Multi-Scale Simulations}
\shortauthors{Hopkins \&\ Most}

\begin{document}

\title{\vspace{-0.8cm}Time-Dilation Methods for Extreme Multiscale Timestepping Problems\vspace{-1.5cm}}

\orcidauthor{0000-0003-3729-1684}{Philip F.~Hopkins}{1,*}
\orcidauthor{0000-0002-0491-1210}{Elias R.~Most}{1}
\affiliation{$^{1}$TAPIR, Mailcode 350-17, California Institute of Technology, Pasadena, CA 91125, USA}

\thanks{$^*$E-mail: \href{mailto:phopkins@caltech.edu}{phopkins@caltech.edu}},

\begin{abstract}
Many astrophysical simulations involve extreme dynamic range of timescales around ``special points'' in the spacetime domain (e.g.\ simulations of black holes, or stars, star formation, planets, disks, galaxies, shocks, mixing interfaces), where processes on small scales couple strongly to those on large scales. Adaptive resolution, multi-physics, and hybrid numerical methods have enabled tremendous progress on the spatial, physics, and numerical challenges of bridging these scales. But often the limiter for following the longer timescales of global evolution and/or small-large scale coupling is the extremely short numerical timestep required in some subdomains (which leads to their dominating the simulation cost). Recently several approaches have been developed for tackling this in problems where the short timescale solution is quasi-periodically sampled and then somehow projected as an ``effective'' subgrid model over longer timescales (e.g.\ ``zooming in and out''). We generalize these to a family of models where the time evolution is modulated by a variable but continuous in space-and-time dilation/stretch factor $a({\bf x},\,t)$. We show how this extends previous well-studied approaches (including reduced-speed-of-light and binary orbital dynamics methods), and ensures that the system comes to the correct local steady-state solutions, and derive criteria that the dilation factor must obey (and associated timestep \&\ resolution criteria) in order to ensure good behavior. We also present a variety of further generalizations of the method to different physics or coupling scales. Compared to previous approaches, this method makes it possible to avoid imprinting arbitrary scales where there is no clear scale-separation, and couples equally well to Lagrangian or Eulerian methods. It is flexible and easily-implemented and we demonstrate its validity (and limitations) in test problems. We discuss the relationship between these methods and physical time dilation in general relativistic magnetohydrodynamics, and use this to inform the behavior of certain source terms that arise in conservative formulations. We demonstrate how this can be used to obtain effective speedup factors exceeding $\gtrsim 10^{4}$ in multiphysics simulations. 
\end{abstract}


\maketitle

\section{Introduction}
\label{sec:intro}

Extreme multi-scale problems define much of the need for numerical simulations in astrophysics (and many other natural sciences). In some contexts -- e.g.\ turbulence, or thermo-chemistry -- every element in volume or mass or time is comparably multi-scale. But in many {astrophysical} applications of interest, there are ``special'' regions of spacetime or subdomains of interest: e.g.\ black holes, stars, planets, galaxies, shocks, clouds, etc. In many subfields, there has been an increasing push towards simulations using super-Lagrangian refinement techniques to ``zoom in'' around some $\mathcal{O}(1)$ number of these special regions and simultaneously follow the dynamics on these scales from very large scales. One specific example is in black hole (BH) feeding, evolution, and ``feedback'' (from radiation, jets, and winds). Recently, novel methods have been independently developed by multiple groups to study this problem from scales of the circum/inter-galactic medium at $\sim$\,Mpc (where gas that is accreted by BHs ultimately originates, and where ``feedback'' from BHs ultimately has strong effects), down to scales of order the BH horizon size at $\sim$\,au \citep[see][]{cho:2023.multiscale.accretion.sims.bondi.inflow.model,cho:2024.multizone.grmhd.sims.bondi.flow.lowaccrate,guo:2024.fluxfrozen.disks.lowmdot.ellipticals,hopkins:superzoom.disk,hopkins:superzoom.overview,hopkins:superzoom.imf,hopkins:superzoom.agn.disks.to.isco.with.gizmo.rad.thermochemical.properties.nlte.multiphase.resolution.studies,kaaz:2024.hamr.forged.fire.zoom.to.grmhd.magnetized.disks}. Another example is simulations of first stars, where simulations have been refining on gas flows from cosmological scales down to as small as stellar radii for more than two decades \citep[e.g.][]{abel:2002.first.star.sim.zoom.to.sub.au.scales,bromm:2004.first.star.accretion.simulations,turk:2009.popIII.binary.formation.zoom.to.solar.radii.resolution,greif:2012.first.stars.sims.sub.solar.radii,bromm:2013.first.stars.sims.reviews}

There are many challenges spanning such a large dynamic range in scales in numerical simulations: novel refinement, domain decomposition, and parallelization techniques are often required; the important physics and allowed physics approximations on various scales may be different; the optimal numerical methods can be distinct; and the timesteps can vary tremendously. This last {point} has, in practice, often proved the rate-limiter to how far such multiscale simulations can be evolved, {even when hierarchical or locally adaptive timestepping algorithms are employed}. In the examples above, the numerical timestep $\Delta t$ can be as short as $<1\,$second in the most-refined regions (owing to extremely short dynamical/sound crossing/light-crossing times) while evolution timescales exceed $\gtrsim 10^{17}\,$seconds at the largest radii. While methods exist that allow different spatial scales and/or physics to be evolved on much smaller timesteps (see references above and e.g.\ \citealt{pan:2022.accelerating.molecular.dynamics.multi.timestepping,wilhelm:2024.venic.code.operator.splitting.multiphysics}), it is simply not possible, at present, to advance the central regions \textit{so many} timesteps as required to follow global evolution timescales (or even remotely close to them).

The traditional approach to this has almost always been some form of ``stitching'' -- taking models or simulations of different scales, reducing their complexity to a couple of input/output numbers in some simple fitting functions or lookup tables, and inserting this into simulations of smaller/larger scales as a ``sub-grid'' model. Essentially, models of smaller (or larger) scales become special inner (or outer) boundary conditions (around some arbitrary number of special sub-regions or points in the domain (${\bf x}_{i},\,t$), like ``black holes'' or ``stars''), which can be deterministic or statistical \citep[see e.g.][]{rodriguez:2018.cluster.hybrid.techniques}. If the problems of interest had limited dimensionality, clean scale separation, and small and large scales were decoupled, then this approach could work perfectly. But of course the high-dimensionality, lack of scale separation, and strong coupling between small and large scales is precisely what makes many astrophysical problems interesting and the subject of active research. But if small and large scales are coupled, we return to the problem of needing to evolve the smallest scales over the large-scale system dynamical/evolution times, to see how e.g.\ an AGN jet is launched and then couples to the cooling in the galaxy (which then in turn re-couples to the accretion flow by determining how much gas can reach the galaxy center), or protostellar jets/radiation/winds regulate their subsequent accretion and formation of nearby stars.
Another thing that makes some of these problems special is that they have almost self-similar power-law type scalings, which for the most part are universal, e.g., for a given accretion regime.

However, it is often the case that small scales can be treated as in ``quasi-equilibrium'' or \textit{statistical} steady-state, over some large number of local dynamical times. This suggests that one might only need to evolve the system on small scales ``fast enough'' to reach equilibrium, then one can effectively use that for some amount of time as a solution on larger scales, until resampling the small scales. 
This is the basic idea behind various related ``equation-free'' \citep{kevrekidis:2002.equation.free.multiscale.methods.microscale.simulators.in.largescale,kevrekidis:2009.equation.free.approaches.to.highly.multiscale.simulation.problems.review} and ``heterogeneous multiscale methods'' \citep{engquist:2005.heterogeneous.multiscale.methods.hmm.multiscale.problem.approaches,weinan:2007.hmm.methods.review} such as projective integration \citep{gear:2003.projective.integration.methods,tretiak:2022.multiscale.methods.twotimescale.average.and.project} for time integration that have been developed in engineering, physical chemistry, terrestrial and climate sciences \citep[see e.g.][]{weinan:2011.principles.multiscale.problems.simulations.hmm.projective.amr.more}. 
Conceptually similar ideas have also been applied to thermonuclear detonation waves in supernova simulations, where ``burning limiter'' methods effectively broaden the detonation front (whose physical width $\sim 1\,$cm is far below the grid scale $\sim 10^{5}\,$cm) while preserving its internal steady-state structure, achieving effective speedups of order $\sim (10^{5}/10^{-1})^{4} \sim 10^{24}$ \citep{kushnir:2020.burning.limiter.detonation.supernova,kushnir:2020.subchandrasekhar.detonations.convergence}. It is also the central idea behind ``slowdown'' methods used to treat hard binaries in collisional N-body simulations \citep{mikkola.aarseth:1996.slowdown.for.close.binaries.dynamics,roman:2012.regularization.similar.coordinates,wang:2020.slowdown.integrators.for.fewbody.problems,hamers:2020.secular.dynamics.slowdown.methods.nested.binaries,szucs:2023.regularization.methods.orbital.dynamics,rantala:2022.bifrost.bh.integration.in.galaxies}. Alternatively, recently \citet{cho:2024.multizone.grmhd.sims.bondi.flow.lowaccrate} and \citet{guo:2025.grmhd.cyclic.zoom.modeling} developed iterative or cyclic zoom-in methods attempting to leverage this: using fixed-mesh methods with a pre-defined refinement scheme around a single central point (a black hole), they effectively ``zoom in,'' activating or refining (with adaptive mesh refinement; AMR) the central regions, evolve them for a pre-specified amount of time, then de-refine or zoom out but effectively freeze certain parts of the fluxes from the small-scale simulation (e.g.\ outward mass/momentum/energy flux, representing a jet or wind), while advancing a larger annulus in time. This effectively re-samples a pre-specified high-resolution region (or concentric set of regions) at fixed intervals, then forces an inner boundary condition or subgrid model for the larger volumes matched to the discrete results of the high-resolution subdomain at its last active timestep. 

Here, we propose a continuous generalization of these methods, applying a smoothly space-and-time varying time ``dilation'' or ``stretch'' factor $a({\bf x},\,t,\,...)$ to elements in the simulations. This extends the methods of \citet{cho:2024.multizone.grmhd.sims.bondi.flow.lowaccrate} and \citet{guo:2025.grmhd.cyclic.zoom.modeling} (which we show are special cases corresponding to specific choices of $a$), and generalized techniques already developed for ``reduced speed of light'' methods (with radiation, neutrinos, or cosmic rays; \citealt{gnedin.abel.2001:otvet,gnedin:2016.reduced.c.scale.separation.background.vs.time,hopkins:m1.cr.closure,ji:2021.mhd.pic.rsol,deparis:2019.reduced.c.ionization.tests,ocvirk:2019.reduced.c.impacts}) or reduced-wavespeed methods for elliptic problems \citep{hopkins:mhd.gizmo}, as well as some versions of equation-free HMM (e.g. projective integration, above). This allows for continuous adaptivity (for problems where there are no special ``breaks'' or gaps between scales), more flexible approaches to refinement (e.g.\ more complex timestep acceleration and/or multiple ``special'' regions within the global simulation domain), and restoring local conservation (if desired), while coupling naturally to $N$-body solvers and Lagrangian methods as well as Eulerian fluid methods with arbitrary individual timesteps. In \S~\ref{sec:method} we derive the method, corrections needed for conservation and numerical stability, adaptation to moving/adaptive special subvolumes, self-correction methods, and various criteria where it should give converged solutions. In \S~\ref{sec:demo} we consider tests in an idealized problem and applications to the problem of BH accretion, showing that it can reproduce the results of brute-force (standard-timestepping) calculations effectively under these circumstances. We discuss applications, advantages, and disadvantages of these methods in \S~\ref{sec:tradeoffs} and summarize in \S~\ref{sec:summary}.

\section{The Method}
\label{sec:method}

\subsection{Basic Idea}
\label{sec:method:basic}

Writing the evolution equations for the fluid state vector ${\bf U}_{i}$ of a resolution element $i$ (around some position ${\bf x} \approx {\bf x}_{i}$ at time $t = t^{n}$) in some conservative form $D {\bf U}_{i}/D t = \mathcal{F}({\bf U}_{j},\,...)$, the basic idea is to modify this to 
\begin{align}
\label{eqn:method} \frac{D {\bf U}_{i}}{D t} \rightarrow \frac{1}{a_{i}}\,\frac{D {\bf U}_{i}}{D t} = \mathcal{F}({\bf U}_{j},\,...)
\end{align}
where
\begin{align}
a_{i} \equiv a({\bf x}_{i},\,t = t^{(n)},\, {\bf U}_{i},\,...) 
\end{align}
is a dimensionless dilation/scale/stretch factor. Note if higher-order time derivatives are explicitly used, one should correspondingly replace $D^{m}/D t^{m} \rightarrow a_{i}^{-m} D^{m}/D t^{m}$, for consistency. This is directly analogous to what is done in reduced speed of light (RSL) methods for radiation/neutrino/cosmic ray dynamics in many different subfields \citep[see e.g.][]{gnedin.abel.2001:otvet,skinner:2013.athena.reduced.c.implementation,gnedin:2016.reduced.c.scale.separation.background.vs.time,hopkins:m1.cr.closure,ji:2021.mhd.pic.rsol,deparis:2019.reduced.c.ionization.tests,ocvirk:2019.reduced.c.impacts}, where there $a_{i} = \tilde{c}_{i}/c$ in terms of the RSL $\tilde{c}_{i}$ (here more accurately akin to ``variable reduced speed of light'' methods; see e.g.\ \citealt{katz:2017.amr.variable.reduced.c.implementation.amr,rosdahl:2018.sphinx.sims.binaries.reduced.c.variable,chan:2024.variable.reduced.c.sims}, or similarly variable damping-wave speed methods developed for numerical divergence-cleaning or other elliptic problems; \citealt{hopkins:mhd.gizmo}). It is also directly analogous to the ``slowdown factor'' applied to hard binaries in regularized N-body simulations \citep{mikkola.aarseth:1996.slowdown.for.close.binaries.dynamics,wang:2020.slowdown.integrators.for.fewbody.problems}.

Just like those methods, this means the definition of $a$ has no effect on local steady-state solutions (i.e.\ it is negligible when $D_{t}{\bf U}_{i}$ is small). Because this effectively slows down evolution the state vector by a factor $a_{i}$, it allows us to formally advance the system by a larger timestep $\Delta t \rightarrow \Delta t a_{i}^{-1}$, whatever the natural timestep criterion would have been. Essentially we just extend this approximation to \textit{all} evolution terms, not just to one subset of the ${\bf U}_{i}$ (e.g.\ the radiation intensity or cleaning-wave amplitude).

We stress that when $a \ne 1$, the system being integrated is \textit{not} the original equations of motion --- the effective evolution equations are modified by the dilation factor, and so the transient, dynamical behavior of the system will generally differ from the undilated case. The method is designed to preserve the correct local steady-state solutions (where $D_{t}{\bf U} \rightarrow 0$), but the path taken to reach those solutions, and any time-dependent or statistically-fluctuating behavior, will be altered by the dilation. This is by construction --- it is the reason the method produces a speedup --- and is directly analogous to the well-known behavior of RSL methods, where transient dynamics differ but the correct steady-state is recovered \citep{skinner:2013.athena.reduced.c.implementation,gnedin:2016.reduced.c.scale.separation.background.vs.time}. We demonstrate this explicitly in the test problems of \S~\ref{sec:demo}, which include both steady-state and non-equilibrium cases.

\begin{figure*}
	\centering
	\includegraphics[width=0.9\textwidth]{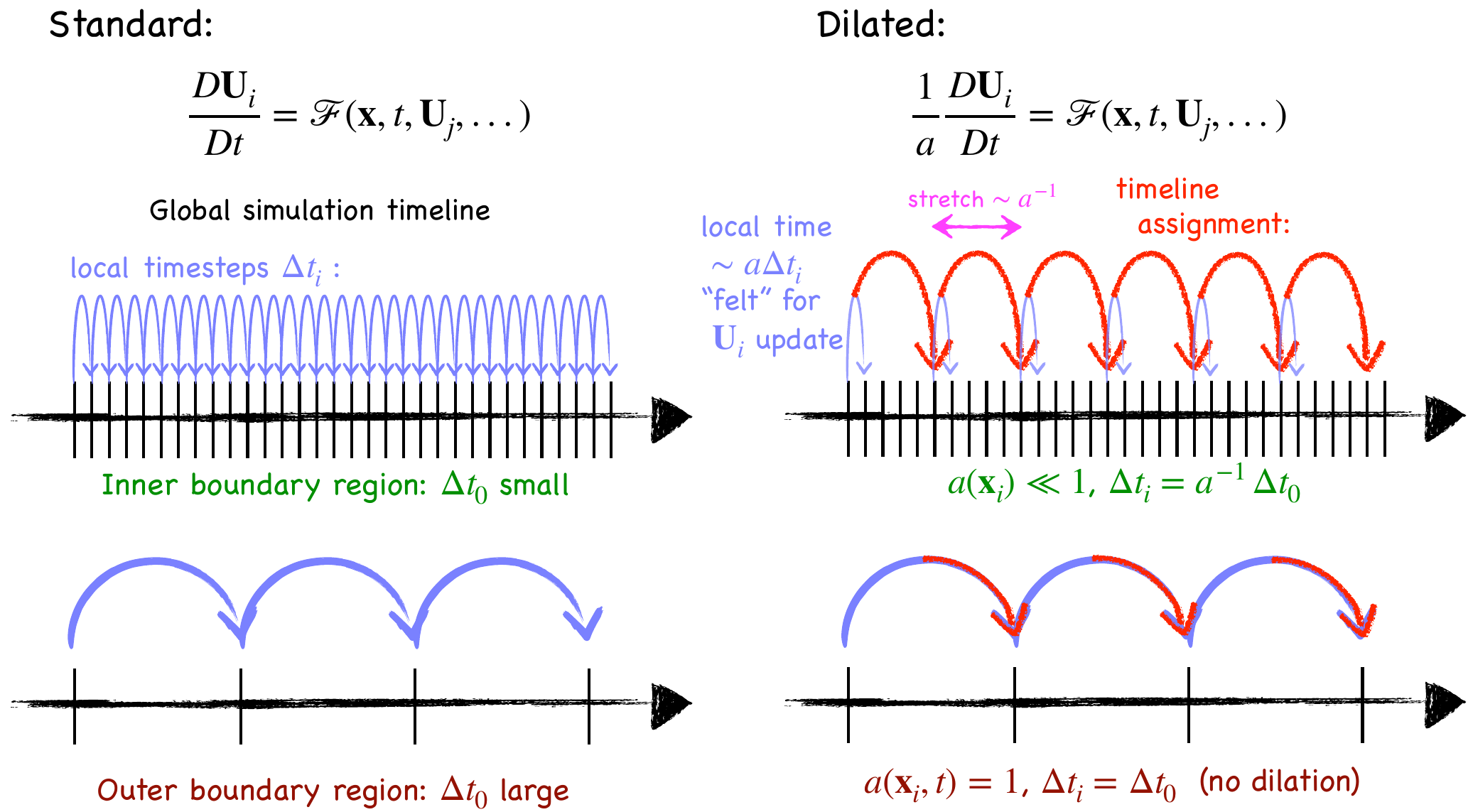} 
	\caption{Heuristic illustration of the method proposed here (\S~\ref{sec:timeline}).
	\textit{Left:} Standard evolution. Variables ${\bf U}_{i}$ are updated according to some time update $\mathcal{F}_{i}$, with 
	a typical CFL-limited $\Delta t_{i} = [\Delta t_{i}]_{0}$, which here is small in some ``inner'' zone and large in some ``outer'' zone. 
	We illustrate how cells in those zones would be advanced along the global timeline in parallel.
	\textit{Right:} Evolving the dilated Eq.~\ref{eqn:method}.
	All fluxes/updates are ``slowed down'' by $a = a({\bf x},\,t)$, allowing for a larger CFL-limited timestep $\Delta t_{i} = a_{i}^{-1} [\Delta t_{i}]_{0}$. 
	Equivalently, on a global timeline update of $\Delta t_{i} = a_{i}^{-1} [\Delta t_{i}]_{0}$, cell $i$ experiences an effective timestep $a_{i} \Delta t_{i}$ in its frame, which is used to calculate the update to ${\bf U}_{i}$, while the cell is moved along the global timeline by $\Delta t_{i}$ (effectively ``stretching'' each timestep by a factor $a_{i}^{-1}$ on the timeline). The outer boundary cells have $a = 1$ so experience no dilation. Inner cells still take smaller updates on the global timeline (obeying conditions in \S~\ref{sec:criteria}), compared to outer cells, so they can respond to secular evolution or continuously changing information. 
	\label{fig:timeline}}
\end{figure*}

\subsection{Implementation: General Concept and Advancement along the Timeline}
\label{sec:timeline}

It is straightforward to implement this in any numerical simulation code (assuming it is capable of individual timestepping of elements $i$), modifying the relevant evolution equations. In Appendix~\ref{sec:lagrangian}, we show that similar to RSL methods one can freely choose for $D_{t} \rightarrow \partial_{t}$ at fixed position (relevant in Eulerian codes with fixed cell position) or $D_{t} \rightarrow {\rm d}_{t}$ (relevant in Lagrangian codes, moving with the cell). And in Appendix~\ref{sec:divergence} we note how this is compatible with standard methods for maintaining divergence constraints (e.g.\ for incompressible fluid dynamics or magnetohydrodynamics).

An alternative and (in many cases) particular convenient way of implementing this is to recognize that if we modify the evolution equations by $D_{t}{\bf U}_{i} \rightarrow a_{i}\, [D_{t}{\bf U}_{i}]_{0}$ and take a larger timestep by $\Delta t_{i} \rightarrow a_{i}^{-1} [\Delta t_{i}]_{0}$, then in any update scheme where we have ${\bf U}_{i}^{n+1} = \hat{\mathcal{F}}_{i} \Delta t \rightarrow a_{i} [\hat{\mathcal{F}}_{i}]_{0} a_{i}^{-1} [\Delta t_{i}]_{0} =  [\hat{\mathcal{F}}_{i}]_{0} [\Delta t_{i}]_{0}$ (where $\hat{\mathcal{F}}_{i]}$ is the effective value of $\mathcal{F}_{i}$ used for the update, which notationally allows for e.g.\ explicit kick-drift-kick or Runge-Kutta type schemes or implicit updates). This means we have just updated the state ${\bf U}_{i}$ as if it took its ``normal'' (smaller) timestep $[\Delta t_{i}]_{0}$. But in the global timestep scheme, we have ``advanced'' it a larger step forward along the timeline 
\begin{align}
\Delta t_{i} = \Delta t_{i}^{\rm timeline} = a_{i}^{-1} [\Delta t_{i}]_{0} = a_{i}^{-1} \Delta t_{i}^{\rm physical}
\end{align}

Fig.~\ref{fig:timeline} illustrates this. Along an integer timeline $N_{t} \equiv 0,\,1,\,...n-1,\,n,\,n+1,\,... $ (essentially, a checklist of tasks/updates corresponding to some time units $t^{(0)},\,t^{(1)},\,... t^{(n-1)},\,t^{(n)},\,t^{(n+1)},\,...$), a step $\Delta t_{i}$ corresponds to evolving the local physical variables in an element $i$ as if their timestep was the ``normal'' timestep allowed by the usual timestep (e.g.\ CFL-type) conditions $[\Delta t_{i}]_{0}$, but advancing the cell some discrete $\Delta n$ such that $\Delta t_{i} \equiv t^{(n + \Delta n)} - t^{(n)}$ along the integer timeline or checklist. The state within each ``normal'' timestep is therefore ``stretched'' over a factor $a_{i}^{-1}$ longer time along the global timeline.  {In the special case of doing this at an inner boundary, this would be identical as keeping $\hat{\mathcal{F}}$ fixed for a number of timesteps, in analogy to the approaches of \citet{cho:2023.multiscale.accretion.sims.bondi.inflow.model,cho:2024.multizone.grmhd.sims.bondi.flow.lowaccrate} and \citet{guo:2025.grmhd.cyclic.zoom.modeling}. We note that these earlier methods effectively implement discontinuous scale factors (with $a$ jumping between $0$ and $1$ at discrete zone boundaries and fixed time intervals), and so must introduce explicit boundary conditions at the interfaces between ``active'' and ``inactive'' zones. By contrast, the continuous $a({\bf x},\,t)$ advocated here avoids these artificial interfaces entirely, which removes the need for special boundary treatments, reduces imprinting of arbitrary scales, and ensures the smoothness criteria (\S~\ref{sec:criteria}) are satisfied everywhere. As we discuss in \S~\ref{sec:tradeoffs:cyclic}, discontinuous approaches can be recovered as a limiting case.}

\subsection{Key Criteria for Good Behavior}
\label{sec:criteria}

There are several basic criteria that the stretch factor $a$ must satisfy in order to ensure desired behavior {(and numerical stability)} in simulations, which we enumerate here.
\begin{enumerate}

\item It must be positive-definite with 
\begin{align}
0 < a \le 1\ .
\end{align}
 Obviously $a<0$ gives the opposite of physical behavior, and $a=0$ is undefined, while $a > 1$ is both counterproductive (requiring more work than standard advancement per unit time) and acausal (it allows cells to move ahead of the global timeline).

\item Outside of proximity to special sub-domains in space+time, 
\begin{align}
a({\rm outside}) \rightarrow 1 \ , 
\end{align}
recovering the standard numerical scheme. This ensures the standard convergence conditions are respected and one is evolving the true equations on these scales.\footnote{Simply evolving $a=1/2$ everywhere, for example, is just the equivalent of taking the final time of a standard simulation run to $t_{f}$ and relabelling this $2\,t_{f}$, and adds no information.}

\item The factor $a$ should vary sufficiently smoothly in space and time, so that we can treat it as constant over a timestep and cell size/spatial step size, or quantitatively
\begin{align}
\label{eqn:smoothness.x}\left |\nabla \ln{a}\right  | &\ll \frac{1}{\Delta x} \ , \\
\label{eqn:smoothness.t}\left |\partial_{t} \ln{a} \right | & \ll \frac{1}{\Delta t} \ .
\end{align}
Moreover as we show below this minimizes errors it introduces in fluxes and conservation. 

\item The problem/domain should be such that any sub-domain with $a \ll 1$ can be approximated as in locally-steady-state over timescales $\lesssim a^{-1}\,[\Delta t]_{0}$.

\item To preserve characteristics (hyperbolic/parabolic/elliptic nature) of the equations being integrated, $a$ should not depend locally (in e.g.\ the computation of fluxes via a Riemann problem or otherwise) on the primitive variables being evolved in ${\bf U}$.

\item It should be the case that regions/sub-domains which evolve on smaller timescales (require smaller $\Delta t$) are able to still take more timesteps in total (so still have smaller timesteps). In other words, if $[\Delta t]_{0}^{\Omega_{1}} \ll [\Delta t]_{0}^{\Omega_{2}}$ in causally-connected sub-domains $\Omega_{1}$ and $\Omega_{2}$, then 
\begin{align}
a_{2}\,\frac{[\Delta t]_{0}^{\Omega_{1}}}{[\Delta t]_{0}^{\Omega_{2}} }< a_{1} \le 1\ , 
\end{align} 
so 
\begin{align}
\Delta t_{\Omega_{1}} < \Delta t_{\Omega_{2}}\ \ \ \ ({\rm if}\ [\Delta t]_{0}^{\Omega_{1}} \ll [\Delta t]_{0}^{\Omega_{2}}) \ .
\end{align}
 Otherwise, these cannot plausibly come into local equilibrium/steady-state (required for the overall validity of the method) on the timescales of the slower-evolving domain.

\end{enumerate}

In the tests we present here (\S~\ref{sec:demo}), we consider $a=a({\bf x})$ (or equivalently $a=a(r)$) for simplicity, as the spatial dependence most directly maps to the radial timestep hierarchy in our target applications. However, all of the derivations above and in \S~\ref{sec:conservative}-\ref{sec:self.valid} hold for general $a=a({\bf x},\,t)$, and indeed $a$ can in principle depend on time explicitly (e.g.\ through the scheduled or adaptive de-dilation schemes in \S~\ref{sec:self.valid}), or implicitly through quantities like the local dynamical time or Mach number. A purely time-dependent $a=a(t)$ (uniform in space) would slow down the entire simulation uniformly and is therefore not useful for multi-scale problems --- the advantage of the method comes specifically from the spatial variation of $a$, which allows different regions to advance at different effective rates. The general $a({\bf x},\,t)$ case with explicit time dependence is most naturally realized through the de-dilation schemes in \S~\ref{sec:self.valid}, where $a$ periodically reverts toward unity to resample the true dynamics.

\subsection{Conservative Forms and Implementation of Flux/Source Terms}
\label{sec:conservative}

Consider a traditional conservative-form equation for some ${\bf U}$, i.e.\ $\mathcal{F} \rightarrow -\nabla \cdot {\bf F} + {\bf S}$ in terms of some flux ${\bf F}$ and source terms ${\bf S}$, modified per Eq.~\ref{eqn:method}: 
\begin{align}
\nonumber \frac{1}{a}\frac{D {\bf U}}{D t} + \nabla \cdot {\bf F} = {\bf S} \ ,\\
\label{eqn:method.base}  \frac{D {\bf U}}{D t} = - a \nabla \cdot {\bf F} + a {\bf S} \ .
\end{align}
Some simple algebra allows us to rearrange this as:
\begin{align}
\nonumber \frac{D {\bf U}}{D t} + \nabla \cdot \left( a {\bf F} \right) &= a {\bf S} + {\bf F} \cdot \nabla a \equiv \tilde{\bf S} \ , \\
\label{eqn:hyper}  \frac{D {\bf U}}{D t} + \nabla \cdot  \tilde{\bf F}  &= \tilde{\bf S} \ , 
\end{align}
with $\tilde{\bf F} \equiv a {\bf F}$ and $\tilde{\bf S} \equiv a {\bf S} + \tilde{\bf F} \cdot \nabla \ln{a}$ as effective flux+source terms. 
Thus the modified equation preserves the hyperbolic/parabolic/elliptic character of the original equations, so long as $\ln{a}$ does not depend explicitly on the primitive variables ${\bf U}$. 

Per \S~\ref{sec:timeline}, the form of Eq.~\ref{eqn:method.base} can be easily implemented by rescaling the time-derivatives from various numerical fluxes, or rescaling the timesteps. But there are also applications where the form of Eq.~\ref{eqn:hyper} may be advantageous, depending on how the numerical implementation treats fluxes and source terms (it requires no modification of the timestepping scheme, in that case, only modifying the inputs to fluxes and source terms sent to some solver). As we discuss in Appendix~\ref{sec:gr}, this makes more explicit the connection to GRMHD as well. 

This also makes the requirement for $a$ to be slowly-varying in space, $| \Delta x \, \nabla \ln{a}| \ll 1$ (\S~\ref{sec:criteria}) more clear. If $a$ had structure on scales small compared to some effective resolution $\Delta x$, then so would $\tilde{\bf S}$ and $\tilde{\bf F}$, but the other terms in ${\bf F}$ and cell areas/reconstructions/gradients would not be able to represent this at the order of integration, so the reconstruction of different terms in Eq.~\ref{eqn:hyper} would not be consistent.

\subsubsection{Interpretation of Conserved-Variable Source/Sink Terms}
\label{sec:source.sink}

Because we adopt a convention such that ${\bf U}$ represents the same primitive variables which would be obtained in the un-dilated ($a=1$) solution, we see that when $a\ne1$ we must modify our interpretation or definition of some conserved quantities. For example, taking the Eulerian continuity equation ${\bf U}\rightarrow \rho$, ${\bf F}\rightarrow \rho\,{\bf u}$, ${\bf S}\rightarrow 0$, we immediately see that the total mass flux through an infinitesimally thin surface is $\dot{M} \rightarrow \oint a \rho {\bf u} \cdot d{\bf A}$. 

This is precisely the desired behavior. Consider, for illustration, a spherical accretion problem like Bondi accretion. The steady-state that is recovered by Eq.~\ref{eqn:method.base} (when $D_{t}{\bf U}\rightarrow 0$), is independent of $a$ in terms of the primitive variables $\rho({\bf x},\,t)$, ${\bf v}({\bf x},\,t)$, etc. The accretion rate $\dot{M}$ defined by these variables, $\dot{M} \equiv -4\pi\,r^{2} \rho(r)\,v_{r}(r)$  is therefore preserved. But this will differ from the $\dot{M}_{\rm s}$ traditionally defined by a sink particle or discrete conservative measurement at an explicitly-defined inner boundary on the grid -- i.e.\ the sum of the explicit cell masses of every cell which crosses the inner boundary $\Omega_{i}$, $\dot{M}_{\rm s} \equiv -\oint_{\Omega_{i}} \rho({\bf x},\,t)\,{\bf v}({\bf x},\,t) \cdot d{\bf A} \approx a_{i} \dot{M}$, where $a_{i}$ is the value $a({\bf x},\,t)$ at the inner boundary $\Omega_{i}$. This should be the case -- $\dot{M}_{\rm s} \equiv D_{t} M_{\rm s}$ is an evolution equation like any other, defined at $\Omega_{i}$ where $a=a_{i}$, so should be dilated in our method. This emphasizes that one must account for $a$ in defining conserved variables, as above.
Note the above assumes that there is either a constant $a_{i}(t)$ on the boundary, or if not one should use an appropriately boundary-averaged value that satisfies the integral above.

Alternatively, in some implementations where e.g.\ the mass of a central sink particle is evolved for long timescales explicitly in the code (e.g.\ a central black hole or star or planet), and evolves slowly compared to the dynamics on the smallest scales, one could ``correct'' for this by increasing the mass of the central sink $M_{\rm s}$ by an augment $\Delta m_{j}/a_{j}$ each time a discrete cell/particle of mass $\Delta m_{j}$ is accreted (with $a=a_{j}$ at the spacetime position of accretion). This would preserve the traditional evolution of $M_{\rm s}$. Numerically, this is identical to the ``fast-slow'' separation of $a$ we describe in \S~\ref{sec:fast.slow:time}-\ref{sec:fast.slow:phys} below -- we are operator splitting the evolution of $M_{\rm s}$, $D_{t} M_{\rm s}$ (which is ``slow'') so it is not dilated, while dilating the dynamical equations otherwise. 

\subsubsection{Short-Range Interactions and Conserved-Variable Exchange}
\label{sec:conservative:short}

It is often the case in conservative Godunov-type methods that global conservation laws are promoted to local conservation laws via exchange of a flux of conserved quantities between neighboring cells. Eq.~\ref{eqn:hyper} makes it clear we can easily do the same, e.g.\ taking $\int_{\Omega_{i}} D_{t} {\bf U} d^{3}{\bf x} =  -\int_{\Omega_{i}} \nabla \cdot \tilde{\bf F} d^{3} {\bf x} = -\oint_{\partial \Omega_{i}} \tilde{\bf F} \cdot d{\bf A} \approx -\sum_{j} \tilde{\bf F}^{\ast}_{ij} {\bf A}_{ij}$. Now $a \rightarrow a^{\ast}$ is included in $\tilde{\bf F}^{\ast} = (a {\bf F})^{\ast}$ but this can be reconstructed by any interpolation method {(with the correct choice depending on the amount of spatial smoothness of $a$)} or defined exactly at the face locations. With Eq.~\ref{eqn:hyper}, a pair $ij$ of neighbor cells with differing $a_{i} \ne a_{j}$, on the same timestep $\Delta t$, exchange an antisymmetric (equal-and-opposite) conserved quantity ${\bf A}_{ij} \cdot \tilde{\bf F}_{ij}^{\ast}$, but then will receive un-equal updates through their source terms $\tilde{\bf S} = \tilde{\bf F}\cdot \nabla \ln a$, assuming they have non-overlapping volumes. With Eq.~\ref{eqn:method.base}, the initial equal-and-opposite flux ${\bf A}_{ij} \cdot {\bf F}_{ij}^{\ast}$ would be computed, but then the actual flux used for each cell in the kick operations would be this multiplied by the unequal $a_{i}$ or $a_{j}$, because $a$ appears there outside the gradient. This is of course the intended behavior: if $a_{i} < a_{j}$, evolution for $i$ is slowed down more than for $j$, so its rate-of-change is more suppressed.

If one wished to remove this, to restore the exact traditional ($a=1$) definition of locally conserved variables (exact antisymmetry on the same timestep between neighbors), then one could (in principle) remove the $\tilde{\bf S}=\tilde{\bf F}\cdot \nabla \ln a$ term in Eq.~\ref{eqn:hyper}. This is equivalent to adding a source term ${\bf S}^{\prime} \equiv -{\bf F} \cdot \nabla a$ to Eqs.~\ref{eqn:method.base}-\ref{eqn:hyper}. But from Eq.~\ref{eqn:method.base} we immediately see that this means the steady-state solutions ($D_{t} {\bf U} \rightarrow \mathbf{0}$) are modified from the true solutions. Consider e.g.\ the trivial case of a hydrostatic, stationary, homogeneous (constant-pressure) medium (${\bf U}=\rho {\bf u}$, ${\bf F} = P = P_{0}$, ${\bf S}=0$), with some $a=a(x)$ along one axis: this produces a spurious force $=P\,\partial_{x} a \hat{x}$ and the steady-state solution requires an opposing pressure gradient $a \partial_{x} P = -P\partial_{x} a$ or $P \propto a^{-1}$. Requirement (3) in \S~\ref{sec:criteria} should help ensure these gradient corrections are always relatively small compared to the physical terms, but they are nonetheless spurious. In \S~\ref{sec:gr} we show that this physically corresponds to gravitational redshifts in GR -- but obviously those are not physical here.

\subsubsection{Long-Range Interactions}
\label{sec:conservative:long}

It is trivial to apply Eqs.~\ref{eqn:method.base}-\ref{eqn:hyper} to long-range forces like gravity, as well, and more generally to elliptic problems where contributions can (in principle) come instantaneously from the entire domain. The most popular methods for Newtonian gravity put it inside the source term, ${\bf S} \rightarrow \rho \nabla \Phi$ (for ${\bf U}=\rho{\bf u}$), where $\nabla \Phi$ can be computed analytically or (for self-gravity) via standard tree ($\nabla \Phi_{i} = \sum_{j} G\,m_{j} ({\bf x}_{j} - {\bf x}_{i}) / |{\bf x}_{j} - {\bf x}_{i}|^{3}$), multipole, for Fourier/particle-mesh methods. The appropriate accelerations then simply need to be multiplied by $a_{i}$ for second-order methods, for higher-order Runge-Kutta methods one can account for the derivatives of $a_{i}$ over the cell/particle displacement in space+time to whatever order is needed.

Again it is obvious that this means in an equal timestep, the pairwise momentum change of two distance particles $ij$ (in, say, an N-body method) with $a_{i} \ll a_{j}$ will not be equal-and-opposite, because $i$ is dilated more strongly than $j$. If one wished to preserve manifest pairwise momentum conservation in the traditional sense then one would be forced to adopt a rescaling of the gravitational acceleration seen by $i$ and $j$ respectively by powers of $a_{j}/a_{i}$. But for $a=a({\bf x},\,t)$, this is the same as changing the long-range force law, which would lead to qualitatively different dynamics in steady-state.

\subsection{Wakeup \&\ Timestep Criteria}
\label{sec:wakeup}

In any method with individual timestepping, it is critical to ensure that neighbor steps do not differ too extremely, and this is especially important in Lagrangian methods where otherwise unphysical effects can occur (e.g.\ fast-moving outflows moving ``through'' an inactive element on a longer timestep before it completes its timestep and ``activates''). This is a well-known issue and standard methods following e.g.\ \citet{saitoh.makino:2009.timestep.limiter} and \citet{durier:2012.timestep.limiter} are sufficient to address it (and already widely-implemented in many codes; see \citealt{hopkins:lagrangian.pressure.sph,hopkins:gizmo,springel:arepo,hubber:gandalf.gizmo.methods}). In short elements $j$ are activated and moved to the smallest active timebin if they interact with a neighbor $i$ whose timestep is sufficiently small relative to theirs. Here we require no fundamental modification to these schemes, only (1) depending on code implementation and notation convention (\S~\ref{sec:timeline}), one should use the correct timestep for comparison, and (2) the awakened cell timestep should correspond to no larger than 
\begin{align}
\Delta t_{j}^{\rm wake} < f_{w}\, \frac{a_{j}}{a_{i}}\,\Delta t_{i}
\end{align}
 with the usual ``safety factor'' $f_{w} \sim 2-4$ per the studies above.

In addition, it is {necessary} to enforce the following timestep conditions: 
\begin{align}
\Delta t_{i} & \le C \frac{a_{i}}{\left |  \partial_{t} a \right |_{i}} \ , \\
\Delta t_{i} & \le C \frac{a_{i}}{v_{\rm sig}  \left |  \nabla a \right |_{i}} \ , 
\end{align}
where $C$ and $v_{\rm sig}$ are the usual Courant factor and signal velocity (slightly less-strict criteria are allowed, but we err on the side of safety here). These ensure the smoothness conditions Eq.~\ref{eqn:smoothness.x}-\ref{eqn:smoothness.t} are met. For a well chosen $a({\bf x},\,t)$, these are redundant (the usual Courant-type conditions will automatically ensure this already), but as such they can avoid pathological cases at almost zero cost.

\subsection{Adding Explicit Fast/Slow Separation}
\label{sec:fast.slow}

\subsubsection{Scale Separation in Time}
\label{sec:fast.slow:time}

It may be the case that there is some explicit scale separation between some ``background'' or slowly-evolving solution, and a faster-evolving solution. In this case we can write the conservative variables as 
\begin{align}
{\bf U} \rightarrow {\bf U}_{\rm slow} + \delta {\bf U}_{\rm fast} \ , 
\end{align}
and apply the dilation factor just to the ``fast'' evolving system: 
\begin{align}
\label{eqn:sep} \frac{D {\bf U}_{{\rm slow},\,i}}{D t}  & \rightarrow \frac{D {\bf U}_{{\rm slow},\,i}}{D t} \ , \\ 
\nonumber \frac{D \delta {\bf U}_{{\rm fast},\,i}}{D t} &\rightarrow \frac{1}{a_{i}}\frac{D \delta {\bf U}_{{\rm fast},\,i}}{D t} \ .
\end{align}
This allows for flexible application where desired, \textit{if} a clear scale separation exists. 

It is easy to come up with examples. For example, returning to the RSL analogy, it is common in cosmological reionization simulations to split the ionizing radiation intensity equation $D_{t} I_{\nu}$ by decomposing $U = I_{\nu}$ into $U_{\rm slow} = \langle I_{\nu} \rangle $ (the cosmic mean background) and $\delta U_{\rm fast} = \delta I_{\nu}$ (the fluctuations from said background), where the slow term can be trivially evolved analytically as a function of redshift, while the fast term is the term explicitly evolved by the radiation hydrodynamics \citep[see discussion in][]{gnedin:2016.reduced.c.scale.separation.background.vs.time}. One could immediately and easily generalize this to include all cosmological mean terms (e.g.\ the cosmic mean density evolution, Hubble flow, etc.) as a slow term in cosmological simulations. 
There are also examples, in e.g.\ evolution of stellar and planetary structure, solid body/elastic/plastic dynamics, and weakly-compressible hydrodynamics, where one separates the mean background properties (e.g.\ evolving $\delta\rho = \rho-\rho_{0}$, rather than $\rho_{0}$; see \citealt{gresho:1990.vortex,tiwari:2003.finite.pointset.method,basic:2022.lagrangian.methods.dealing.with.sharp.edges.novel.method.sharp.boundaries.for.sph.mfm.to.use}). In those methods, it is essentially the same exercise to apply Eq.~\ref{eqn:sep} to the $\delta{\bf U}$ terms while evolving the background secularly (analytically). 

Another example is if the entire simulation occurs in a moving or free-falling frame with some slowly evolving ${\bf v}_{\rm sim}(t)$, ${\bf a}_{\rm sim}(t)$. Then it is straightforward to decompose ${\bf v}({\bf x},\,t) = {\bf v}_{\rm sim}(t) + [{\bf v}({\bf x},\,t) - {\bf v}_{\rm sim}(t)]$, etc. This is functionally equivalent to de-boosting the simulation to the locally free-falling comoving frame before computing any numerical evolution terms.

\subsubsection{Scale Separation in Space}
\label{sec:fast.slow:space}

Fast ``sub-regions'' in spacetime are already accounted for in the default implementation of the model via the dependence $a({\bf x},\,t)$. However, one could imagine forces/time derivatives coming from different spatial scales which would be treated differently -- i.e.\ decomposing 
\begin{align}
\label{eqn:long.short} {\bf U} &\rightarrow {\bf U}_{\rm long} + \delta {\bf U}_{\rm short} \\
\nonumber {D_{t} \delta {\bf U}_{{\rm short},\,i}} &\rightarrow {a^{-1}_{i}}{D_{t} \delta {\bf U}_{{\rm short},\,i}} 
\end{align}
into some kind of ``short'' range (and fast) versus ``long'' range (and slow) terms, akin to the fast/slow decomposition above. 
{We here make the assumption that the short region does not (appreciably) alter the long/slow dynamics.}

A trivial example is if the entire simulation domain is moving under the influence of a uniform external acceleration ${\bf a}_{0}(t)$ and/or initial boost ${\bf v}_{0}(t)$, so $a({\bf x},\,t)$ can be written $a({\bf x} - {\bf x}_{\rm shift},\,t)$ with ${\bf x}_{\rm shift}(t)$ representing this boost+acceleration. One can then operator-split the global shift ${\bf x}\rightarrow {\bf x} + {\bf v}_{0} t + {\bf a}_{0} t^{2}/2$ and acceleration ${\bf v} \rightarrow {\bf v} + {\bf a}_{0} t$, from the dynamically-evolved forces which are modified by $a$. This is just equivalent to transforming to the free-falling, stationary lab frame before calculating any internal dynamics. 

Generalizing this, consider the case of $a({\bf x}-{\bf x}_{i},\,t)$ which decreases around a special point ${\bf x}_{i}$, which itself represents an object that is allowed to dynamically move in the simulation under the influence of long-range forces (say, a black hole or planet or star embedded in some large-scale environment). Note one can have an arbitrary number of ${\bf x}_{i}$ points (numerically each can broadcast its ${\bf x}_{i}$ so all points in the domain know their closest $i$). 
A specific example of particular interest would be simulations of circumbinary disks, following gas with negligible self-gravity around two Keplerian masses in a binary orbit, where one wishes to zoom-in (with dilation) on the mini-disks that form closely-bound to each point mass. {As cautioned above, this requires subleading feedback from the fast to the slow scales. I.e., the masses cannot change on a rapid timescale that would alter the (slow) orbital dynamics.}
If ${\bf x}_{i}(t)$ has some large-scale slow motion (say its global orbital motion in its host galaxy/cloud/disk, or the motion of each of the masses in the binary), we wish to dynamically follow this for all cells within the region on that global timescale, or else (if their displacement is slowed by $a$) they could artificially lag (e.g.\ fall behind their ``host''). 
Provided $a({\bf x}-{\bf x}_{i},\,t)$ only decreases to values $\ll 1$ in a sufficiently small region $|{\bf x}-{\bf x}_{i}| \ll R_{\rm influence}$ interior to which the object $i$ strongly dominates the gravitational dynamics (one is only dilating elements strongly bound to $i$), then this is straightforward: one can calculate the total long-range gravitational force on $i$ as usual, to obtain ${\bf a}^{\rm grav}_{i} = D_{t}{\bf v}_{i}$, and move all the cells within that dilated domain around $i$ with $i$, i.e.\ taking their individual ${\bf a}_{j} = D_{t} {\bf v}_{j} \rightarrow {\bf a}^{\rm grav}_{i} + a_{j}\,[{\bf a}_{j} - {\bf a}_{i}^{\rm grav}]$ and ${\bf v}_{j} = D_{t} {\bf x}_{j} \rightarrow {\bf v}_{i} + a_{j}\,[{\bf v}_{j} - {\bf v}_{i}]$.  In Eq.~\ref{eqn:long.short} this is equivalent to defining 
${\bf U}^{\rm long}_{j} \rightarrow \{ {\bf x}_{i}, \, {\bf v}^{\rm grav}_{i}, \, {\bf 0},\, ...\}$ and $\delta {\bf U}^{\rm short}_{j} \rightarrow  \{ {\bf x} - {\bf x}_{i}, \, {\bf v} - {\bf v}^{\rm grav}_{i}, \, {\bf U}^{\rm other}_{j},\, ...\}$.
Care is needed, however, if $a({\bf x}-{\bf x}_{i},\,t) < 1$ further from the points ${\bf x}_{i}$ where the most naive applications of this could produce spurious motions (e.g.\ one does not wish for gas at $|{\bf x}-{\bf x}_{i}|\rightarrow \infty$ from the binary to artificially oscillate back and forth with the binary orbit). Specific cases like a circumbinary disk can be handled by more careful selection of a taper function for $a$ and/or definitions of ${\bf U}_{\rm long}$, $\delta {\bf U}_{\rm short}$. 

Another specific example of this is the ``slowdown'' method for treating hard binaries in collisional N-body dynamics \citep{mikkola.aarseth:1996.slowdown.for.close.binaries.dynamics}. Specifically there one applies a dilation factor $a_{ij} \equiv a_{ij}(i,\,j,\,t,\,...)$ ($a = 1/\kappa$ in terms of the ``slowdown factor'' $\kappa$ defined therein) to the short-range gravitational forces and motion $D_{t} \delta {\bf U}_{{\rm short}}$ between identified hard binary stellar pairs $i$ and $j$, while the forces $D_{t} {\bf U}_{\rm long}$ with all longer-range $N$-body particles are un-dilated. This leverages the fact that there is a clear scale-separation in the problem between a given hard binary and external perturbers, although subsequent work \citep{roman:2012.regularization.similar.coordinates,wang:2020.slowdown.integrators.for.fewbody.problems,hamers:2020.secular.dynamics.slowdown.methods.nested.binaries,szucs:2023.regularization.methods.orbital.dynamics,rantala:2022.bifrost.bh.integration.in.galaxies} has shown that care is still needed in defining the function $a$ to avoid corrupting the solutions when the scale separation is not so clear (e.g.\ in hierarchical multiples), because $a$ in those methods is applied discretely \textit{only} to the designated binary, not to e.g.\ everything within some volume (so it is more challenging to satisfy our conditions in \S~\ref{sec:criteria}).


It is also possible to imagine more generalized methods which make $a$ a function of separation, for long-range forces. For example for hybrid gravity solvers which treat shorter-spatial-scale gravity via a tree/fast multipole method and long-range forces via particle-mesh \citep{springel:gadget}, the long-range terms are assumed to be more slowly-evolving. But there more care is needed to (1) avoid imprinting artificial scales (e.g.\ the tree-PM dividing range is purely numerical in most cases, so one does not want a strongly discontinuous $a$), and (2) avoid imprinting spurious gradients or forces. For (2) note that this long-range force division is (in its most naive form) equivalent to the pair-dependent $a^{\ast}$ for long-range forces discussed in \S~\ref{sec:conservative:long}, so has the same drawbacks and introduced artifacts.

\subsubsection{Scale Separation in Physics and Expanded ``Reduced Speed of Light''  Schemes}
\label{sec:fast.slow:phys}

Alternatively, one could apply this separation to different \textit{physics}, operator-splitting $D_{t} {\bf U}$, as: 
\begin{align}
\label{eqn:slow.phys} 
\frac{D {\bf U}_{i}}{D t}  & \rightarrow \frac{D {\bf U}_{{\rm slow,\,phys},\,i}}{D t} + \sum_{\alpha} \frac{1}{a^{\alpha}_{i}} \frac{D {\bf U}^{\alpha}_{{\rm fast,\,phys},\,i}}{D t} \ ,
\end{align}
where the $\alpha$ allows for different operator-split sets of variables, each of which in principle has their own dilation factor $a_{i}^{\alpha}$. We then apply Eq.~\ref{eqn:sep} after operator splitting the evolution of the slow and fast variables. 

Upon some reflection, it should be clear that this is precisely the definition of (well-posed) RSL schemes (\S~\ref{sec:method:basic}). 
If we take $\delta {\bf U}_{\rm fast,\,phys}$ to be the radiation variables $e_{{\rm rad},\,\nu}$, $F_{{\rm rad},\,nu}$, $I_{\nu}$; or corresponding neutrino variables; or similar cosmic ray bulk variables (for CR-MHD schemes as \citealt{hopkins:m1.cr.closure}); or individual relativistic particle variables (${\bf x}_{i}$, ${\bf p}_{i}$) or phase-space distribution function $f({\bf x},\,{\bf p},\,t)$ (for MHD-PIC schemes as \citealt{ji:2021.mhd.pic.rsol}); then we recover those respective RSL schemes. 
Indeed, as emphasized by \citet{skinner:2013.athena.reduced.c.implementation,gnedin:2016.reduced.c.scale.separation.background.vs.time,ji:2021.mhd.pic.rsol}, in an RSL scheme one is not actually ``reducing the speed of light'' -- $c$ must appear with its full physical value everywhere (including all pre-factors, fluxes, source terms, physical definitions, etc.) \textit{except} that one writes the physical time derivatives as $c^{-1}\,D_{t} {\bf U} = c^{-1} \mathcal{F}$ then takes $c^{-1}\,D_{t} {\bf U} \rightarrow \hat{c}^{-1}\,D_{t} {\bf U} = a^{-1}\,c^{-1}D_{t} {\bf U}$ with $a \equiv \hat{c}/c$ in terms of the ``RSL'' $\hat{c}$.

It is less immediately obvious but similar if we take the formulation in Appendix~B of \citet{hopkins:mhd.gizmo} for variable-damping-wavespeed formulations of the \citet{dedner:2002.divb.cleaning.scheme} divergence-cleaning scheme, we can see that the variable wavespeed formulation under the action of divergence correction alone is exactly identical to this (replacing all time-derivatives for the cleaning waves with $a^{-1}_{i} \partial_{t}$, where $a_{i} \equiv v_{{\rm wave},\,i}/v_{\rm wave,\,max}$). 

One can imagine numerous other generalizations of this. For example in some problems self-gravity is a very weak, slowly-evolving force (e.g.\ weakly-self-gravitating disk simulations) compared to hydrodynamics and/or advection/external gravity. The opposite can also be true. In some cases magnetic terms evolve much faster than hydrodynamic (common with e.g.\ whistler waves, allowing for Hall MHD; for which analogous physical and numerical attempts to ``cap'' the speed of whistler waves have been discussed in e.g.\ \citealt{amano:2015.whistler.wave.finite.upper.speed.enforcement}).

\subsubsection{Generalized Dilation-Factor Criteria}
\label{sec:fast.slow:crit}

For the operator-split examples above in \S~\ref{sec:fast.slow:time}-\ref{sec:fast.slow:phys}, we can briefly examine the criteria on $a$ given in \S~\ref{sec:criteria}. Criteria (1)-(4) immediately cross-apply to any of these examples. We need to slightly generalize criterion (5), which argued that sub-domains with some $a < 1$ which would ``normally'' require more (smaller) timesteps than other sub-domains must still take more timesteps after dilation (they just do not need to take ``as many more''). Now, instead of $a_{i}$ just varying between spatial-temporal sub-domains $\Omega$ within the simulation, we now allow for $a_{i}$ to vary between \textit{operations} (operator-split equations representing either fast-slow spatial-temporal terms, or different physics). The straightforward generalization is to require that any operator/equation which would have required more (shorter) timesteps to evolve must still require more (shorter) timesteps to evolve (they just do not have to be ``as short''): 
\begin{align}
\Delta t_{\rm fast} = a^{-1} [\Delta t_{\rm fast}]_{0} < \Delta t_{\rm slow} \ . 
\end{align}
Referring back to the RSL example, this is identical to the well-known and well-tested condition that ``the RSL is still faster than other speeds in the problem'' \citep[e.g.][]{deparis:2019.reduced.c.ionization.tests,ocvirk:2019.reduced.c.impacts}: in other words, if $c \gg v_{\rm sig}$, then $\hat{c} = a\,c > v_{\rm sig}$ is still required, such that the ``radiation timestep'' $\Delta t_{\rm rad} = C_{\rm CFL}\, \Delta x/\hat{c} = a^{-1} C_{\rm CFL}  \, \Delta x/ c$ is still smaller than the ``other physics'' timestep $\Delta t_{\rm other} = C_{\rm CFL} \, \Delta x / v_{\rm sig}$.

\subsection{Self-Validation and Correction Schemes}
\label{sec:self.valid}

The dilation approach should work when the system is in statistical steady-state. In principle (if enough about the problem is known), design of appropriate $a({\bf x},\,t)$ can ensure this. However in many cases one may wish to quasi-periodically revert to the ``brute force'' solution to capture non-steady-state phenomena. One can do so in a ``scheduled'' manner, or an adaptive manner, for which we outline schemes below.

\subsubsection{Scheduled De-Dilation}
\label{sec:self.valid:scheduled}

Consider, for example a desired or target $a = a_{0}({\bf x},\,t)$, and define the actually-used $a$ by:
\begin{align}
\label{eqn:dedilation} a \rightarrow a_{0}({\bf x},\,t) + \left[ 1 - a_{0}({\bf x},\,t) \right]\,\mathcal{P}({\bf x},\,t)
\end{align}
where, $\mathcal{P}$ is some quasi-periodic function that switches between $0 \le \mathcal{P} \le 1$. For example, $\mathcal{P}({\bf x},\,t) = |\sin{(\pi\,\phi[{\bf x},\,t])}|^{2\,\ell}$ in terms of an exponent $\ell$ (higher values $\ell \gg 1$ corresponding to ``sharper,'' more punctuated times where $a \rightarrow 1$) and phase function $\phi$ (determining when $a\rightarrow 1$, every integer increase in $\phi$). 
A special case of this (with discontinuous $a$) reduces to the ``cyclic'' approaches employed in \citet{cho:2024.multizone.grmhd.sims.bondi.flow.lowaccrate,guo:2025.grmhd.cyclic.zoom.modeling} -- their schemes are equivalent to taking both $a_{0}({\bf x},\,t)$ and $\mathcal{P}$ as step functions alternating between $0$ (or $\epsilon$ very small) and $1$. 
One can easily design the function $\mathcal{P}$ to correspond to some sensible physical approach -- for example, activating every $N$ global dynamical times, and moving $a \rightarrow 1$ for a few dynamical times at each scale ``outside in'' approaching some central refinement region, then ``inside out'' (or for example by defining $a \sim 1 / (1 + (|{\bf x}-{\bf x}_{0}| / r_{0})^{-\psi})$ around some central point  ${\bf x}_{0}$, with $r_{0}(t)$ some appropriate function of time). 

\subsubsection{Adaptive De-Dilation}
\label{sec:self.valid:adaptive}

More generally, one could trigger ``de-dilation'' ($\mathcal{P}\rightarrow 1$, in Eq.~\ref{eqn:dedilation}) dynamically on-the-fly, with some criterion. 
Recall the methods here should work well when the system is in approximate statistical steady-state, meaning $\langle D_{t}{\bf U} \rangle$ is (in some appropriate ensemble average) is small. This can be computed with some running average, with some threshold defined to trigger a dedilation. The actual dedilation itself can be triggered following Eq.~\ref{eqn:dedilation} with something like the example $\mathcal{P}({\bf x},\,t) = |\sin{(\pi\,\phi[{\bf x},\,t])}|^{2\,\ell}$, starting from phase $\phi=0$ with an appropriately-chosen physical duration. 

To give a specific example, consider the case of $a_{0}({\bf x},\,t) = a_{0}(r) = 1/(1 + (r/r_{0})^{-\psi})$ with $r=|{\bf x}-{\bf x}_{0}|$ around a central Keplerian mass (e.g.\ black hole, planet, star). A natural/characteristic timescale at each $r$ is the dynamical time $t_{\rm char} = t_{\rm dyn} \equiv 1/\Omega = \sqrt{r^{3}/G M} \propto r^{3/2}$, and a natural volume is the volume enclosed in $r$. So one can define $\langle D_{t} {\bf U} \rangle = \sum_{\rm cells} \sum_{\rm time} W(i,\,t)\,D_{t} {\bf U}_{i}(t) / \sum_{\rm cells}\sum_{\rm time} W(i,\,t)$ interior to some radius $r$ over the last $\Delta t_{\rm avg} = t_{\rm dyn}(r)$, where $W$ is some weighted function (say a boxcar or Gaussian kernel in space and time). Then compare $|\langle D_{t} {\bf U} \rangle|$ (for whatever subset ${\bf U}^{\rm check}$ of variables are being compared) to some threshold $C\,U_{\rm char} / t_{\rm dyn}(r)$, where $C$ is an $\mathcal{O}(1)$ CFL-like constant, and $U_{\rm char}$ some characteristic value of ${\bf U}$, so that dedilation is triggered if: 
\begin{align}
\left | \left \langle \frac{D {\bf U}^{\rm check}}{D t} \right \rangle \right | > C\,\frac{U_{\rm char}^{\rm check}}{t_{\rm char}} \ .
\end{align}
One could do this in principle for every variable ${\bf U}$, or just a subset which are most important. For the Keplerian case, for example, one could take ${\bf U}^{\rm check} \rightarrow {\bf v}$, and $U_{\rm char}^{\rm check} \rightarrow v_{K} \equiv \sqrt{G M/r}$.  

An even simpler way of doing this is to note that the average rate-of-change (for an appropriately weighted average) of some conservative variable ${\bf U}$ over finite volume and time is simply the net change in the conserved quantity. Thus every $\Delta t^{\rm check} \sim t_{\rm char}$ (or even after every integer timeline step), one can compute the total of some conserved quantity 
\begin{align}
{\bf Q}^{\rm check} \equiv \sum_{{\rm elements}\,i} ({\bf U}^{\rm check}_{i} V_{i})
\end{align}
(e.g.\ mass/momentum/energy, for ${\bf U}^{\rm check}_{i} = \rho,\,\rho\,{\bf v},\,e$). Then consider the fractional change in that quantity: 
\begin{align}
\frac{t_{\rm char}}{\Delta t^{\rm check}} \frac{\left | {\bf Q}^{\rm check}(t) - {\bf Q}^{\rm check}(t - \Delta t^{\rm check}) \right |}{\left | {\bf Q}^{\rm check}(t) + {\bf Q}^{\rm check}(t - \Delta t^{\rm check}) \right |}  \ ,
\end{align}
and if this exceeds some dimensionless threshold $C$, trigger a dedilation.
So, for example, taking ${\bf U}^{\rm check} \rightarrow \rho$, $\Delta t^{\rm check} = t_{\rm char} = 2\pi\,t_{\rm dyn}$ for our Keplerian example, every orbital time at $r$ ($2\pi\,t_{\rm dyn}(r)$), we check if the fractional change of the total mass inside of $r$ over the past orbital time exceeds a dimensionless threshold $C<1$ which defines our tolerance for rapid changes. If it exceeds this threshold, we trigger a dedilation inside of $r$, raising $a(r^{\prime} < r,\,t)$ to be equal to $a(r)$ (so as to not introduce any artificial discontinuity or inversion of $a$), until the criterion is again satisfied, at which point we re-dilate inside of $r^{\prime} < r$. 

\section{Validation \&\ Examples}
\label{sec:demo}

We now discuss several validation and example problems of these methods. For the specific numerical examples here, we implement this method in the code {\small GIZMO} \citep{hopkins:gizmo,hopkins:mhd.gizmo,hopkins:cg.mhd.gizmo,hopkins:gizmo.diffusion}, a flexible multi-method Lagrangian fluid dynamics+gravity+multiphysics code. We follow the ``stretched timeline'' implementation in \S~\ref{sec:timeline}, which means $a({\bf x},\,t)$ is applied to all Lagrangian time derivatives and operators equally (not singling out specific physics for dilation). Because {\small GIZMO} already allows for arbitrary individual timesteps, this is straightforward. We provide the source code and example implementation for the idealized test problems as part of the public {\small GIZMO} code \citep{hopkins:gizmo.public.release}.\footnote{Available at \gizmourl}

\subsection{Existing Test Problems \&\ Validation}
\label{sec:demo.existing}

First, we recall that since this is a generalization of many existing well-tested methods, that means example problems for ``special cases'' of $a({\bf x},\,t)$ already exist in the literature. 

For example, consider variable RSL implementations. The test problems for these all fall under some $a({\bf x},\,t)$ applied to the radiation physics, but any test problems which \textit{only} evolve the radiation variables are effectively applying a variable $a({\bf x},\,t)$ over the entire domain to all evolved variables, by definition. This means test problems like those in \citet{katz:2017.amr.variable.reduced.c.implementation.amr}, Appendix~A, specifically their ``point source in an optically-thin box'' (testing a coupled set of advection-diffusion equations), their ``point source in a low-density medium'' (similar equations but with non-zero source+sink and scattering rates), and Iliev test 6 variant (expanding I-front) tests all demonstrate that these methods converge rapidly to the correct steady-state solutions. The same is true of the tests in \citet{chan:2024.variable.reduced.c.sims}, including their ``beam/1D propagation'' test, and steady-state ``photon-bounded HII region'' tests. Likewise for the steady-state ``pure-divergence cleaning'' wavespeed test in \citet{hopkins:mhd.gizmo}. There are also a wide range of tests of the slowdown methods for hard binary integration, discussed in \citet{mikkola.aarseth:1996.slowdown.for.close.binaries.dynamics,wang:2020.slowdown.integrators.for.fewbody.problems,hamers:2020.secular.dynamics.slowdown.methods.nested.binaries,rantala:2022.bifrost.bh.integration.in.galaxies}, which demonstrate that applying a time-dilation factor $a$ to the short-range interactions between a hard binary pair successfully recovers the secular evolution of those binaries from long-range forces. 

A more distinct set of tests are presented for the ``cyclic zoom'' methods in both \citet{cho:2023.multiscale.accretion.sims.bondi.inflow.model} and \citet{guo:2025.grmhd.cyclic.zoom.modeling}, both showing variations of the steady-state Bondi-Hoyle accretion problem as validation of their methods. We will consider that as well for completeness.

\subsection{Idealized Test Problems}
\label{sec:demo.test}

We now consider two idealized test problems with different physics and numerical constraints.

\begin{figure}
	\centering
	\includegraphics[width=0.98\columnwidth]{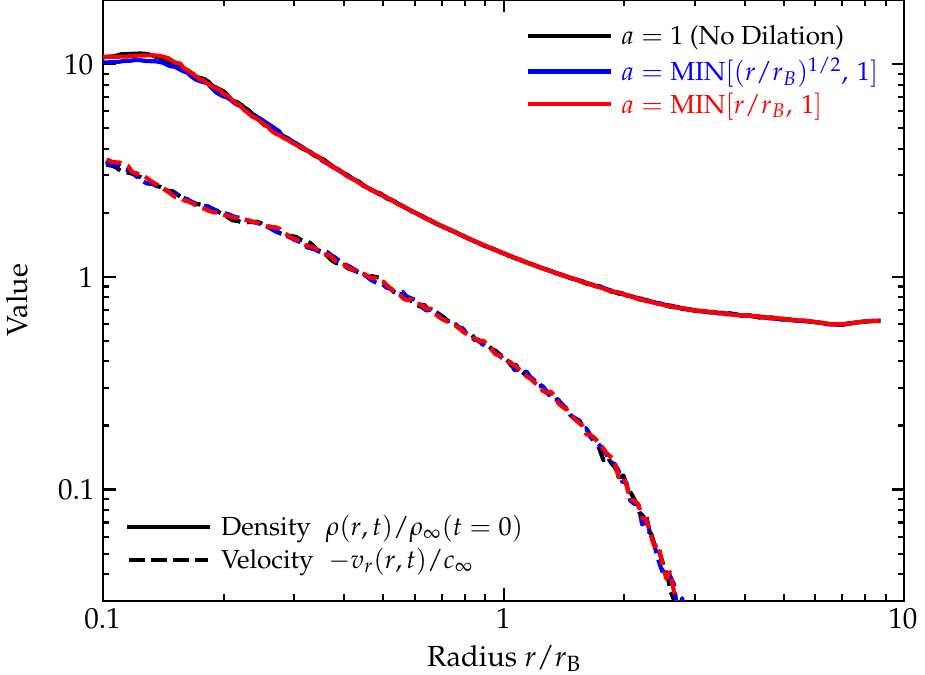} 
	\caption{3D secularly-evolving spherical (Bondi) accretion test (\S~\ref{sec:demo.test:bondi}). We initialize a uniform-density finite-mass periodic box of isothermal gas in a Keplerian, accreting potential, and allow it to evolve. We plot density and inflow velocity profiles at a time where a significant fraction of the box mass has been depleted. The time-dilation methods ($a < 1$) reproduce the solutions of the standard (no dilation, $a=1$) method up to integration errors. They capture the correct Bondi-like steady-state but also the continuous secular evolution as the box mass supply (hence $\rho_{\infty}$) depletes slowly.
	\label{fig:bondi}}
\end{figure} 

\subsubsection{Local Steady-State with External Gravity and Secular Evolution: Spherical (Bondi) Accretion}
\label{sec:demo.test:bondi}

First, consider the case of spherical, isothermal, \citet{1944MNRAS.104..273B} accretion onto a Keplerian point mass in a homogeneous medium. This has high relevance to some applications but simplified physics and numerics. Numerically, we treat gravity as analytic and Keplerian from a point mass at the origin with $G=M=1$: there is no self-gravity. Gas obeys a strict isothermal equation of state and $c_{s}=c_{\infty}=\sqrt{P/\rho}=1$, and is initialized to have uniform density $\rho_{\infty}=1$ throughout the box with a glass cell configuration. The Bondi radius is at $r=r_{B} = 1$, the inner boundary at $r=0.1$ is a pure accretion/inflow/sink boundary, the domain is a periodic cube in 3D of size $L=10$. We evolve for $\sim 40\,r_{B}/c_{\infty}$. This test was run with the meshless finite mass (MFM; \citealt{hopkins:gizmo}) hydrodynamic solver in {\small GIZMO}, with $50^{3}$ initial equal-mass resolution elements, but we verified that other solvers behave similarly.

We compare three simulations in Fig.~\ref{fig:bondi}. First our reference case $a=1$, i.e.\ no dilation. Second a case with $a = {\rm MIN}[(r/r_{B})^{1/2},\,1]$, 
and third a case with $a={\rm MIN}[(r/r_{B}),\,1]$. We have also run a case with $a = 1/(1 + 1/r)$ at $t=0$, where we twice (at one-third and two-thirds of the run time) de-activate the dilation to resample the solution on its ``regular'' timesteps for a time $\Delta t = 0.05$, which gives indistinguishable results. 

We see that the system rapidly comes into quasi-steady-state, with very weak evolution in the $\rho$ and $v_{r}$ and $\dot{M}$ profiles. There is (intentionally) a slow secular evolution term, as the box is finite-mass and only extends to a few $r_{B}$, so is gradually depleted ($\rho_{\infty}$ decreases with time). The methods here recover the $a=1$ solution up to basically the same numerical errors as the standard method. In particular, the time-averaged mass flux $\langle \dot{M}(r) \rangle$ through spherical shells at different radii agrees between the dilated and un-dilated runs to within $\lesssim$\,a few percent, consistent with the numerical integration error at this resolution. But this is not completely trivial: while the methods here should automatically preserve steady-state, here we test whether they can reach that steady-state. Moreover we explicitly set this up to be a problem with a fast-slow time separation, featuring secular evolution of $\rho_{\infty}$ and therefore $\dot{M}_{B}$, which is captured correctly by the method. We also  refer to the unmagnetized and magnetized Bondi tests in \citet{cho:2023.multiscale.accretion.sims.bondi.inflow.model} and \citet{guo:2025.grmhd.cyclic.zoom.modeling}, as further demonstrations for specific examples of $a({\bf x},\,t)$.

Related to \S~\ref{sec:conservative}, the point therein arises about whether we choose to augment the central analytic mass used for gravity as mass is accreted, according to the (time-dilated) mass which flows through the boundary or some corrected version of that. But because we are considering the effective test-particle limit (the classical problem with negligible self-gravity) so our density units are not meaningful, and even if they were $M_{\rm s}$ changes by a small amount over the duration of the simulation (by design), the evolution of the test problem is indistinguishable whichever of these approaches we adopt, and modifying $D_{t} M_{\rm s}$ as we describe (because it is only recorded analytically at the boundary for purposes of the analytic gravity term) is just a one-line bookkeeping change and has no effect on our plotted results.

\begin{figure}
	\centering
	\includegraphics[width=0.97\columnwidth]{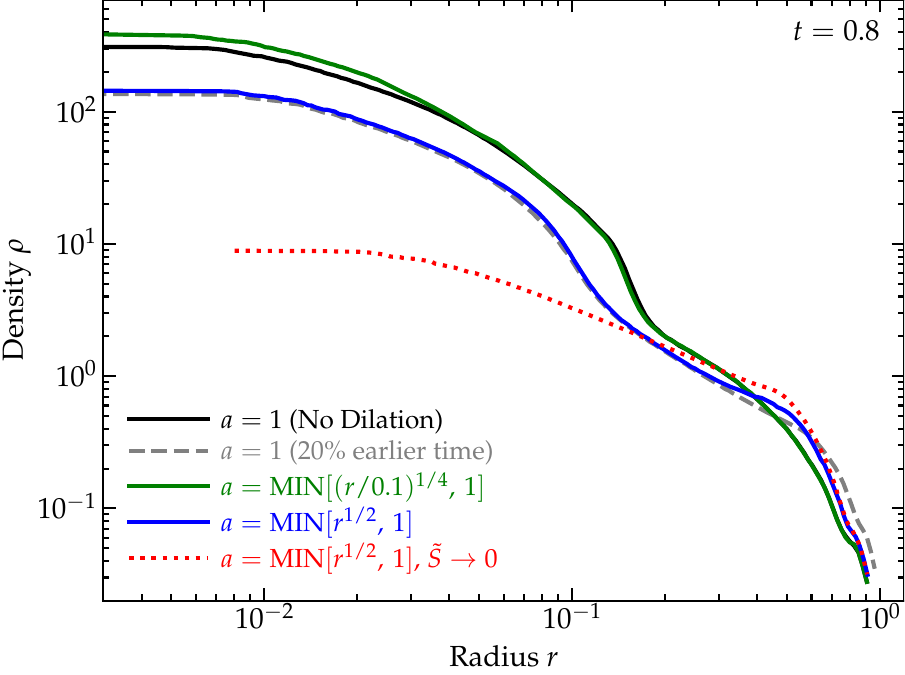} 
	\includegraphics[width=0.97\columnwidth]{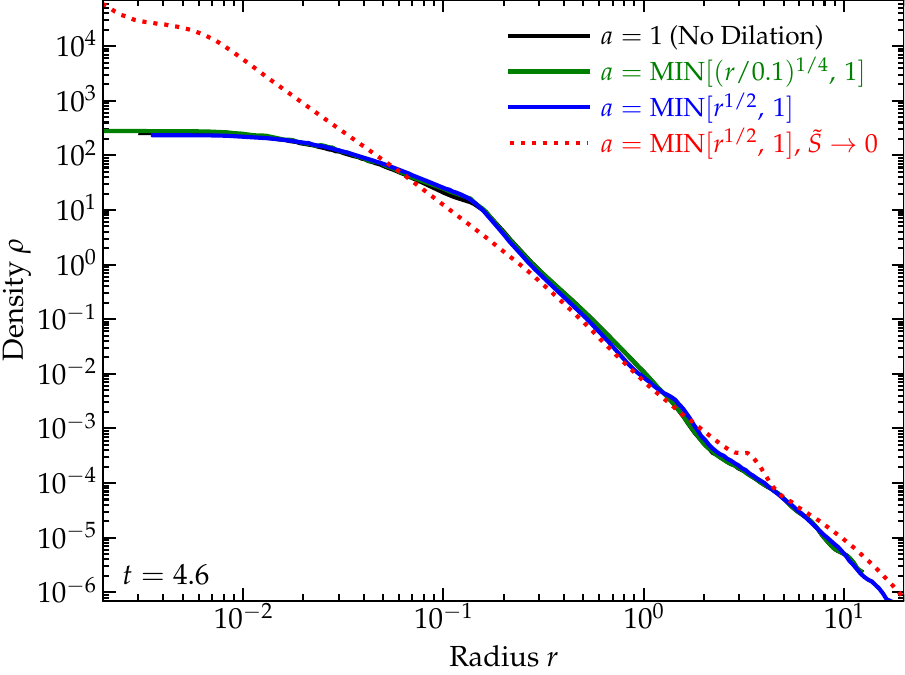} 
	\includegraphics[width=0.98\columnwidth]{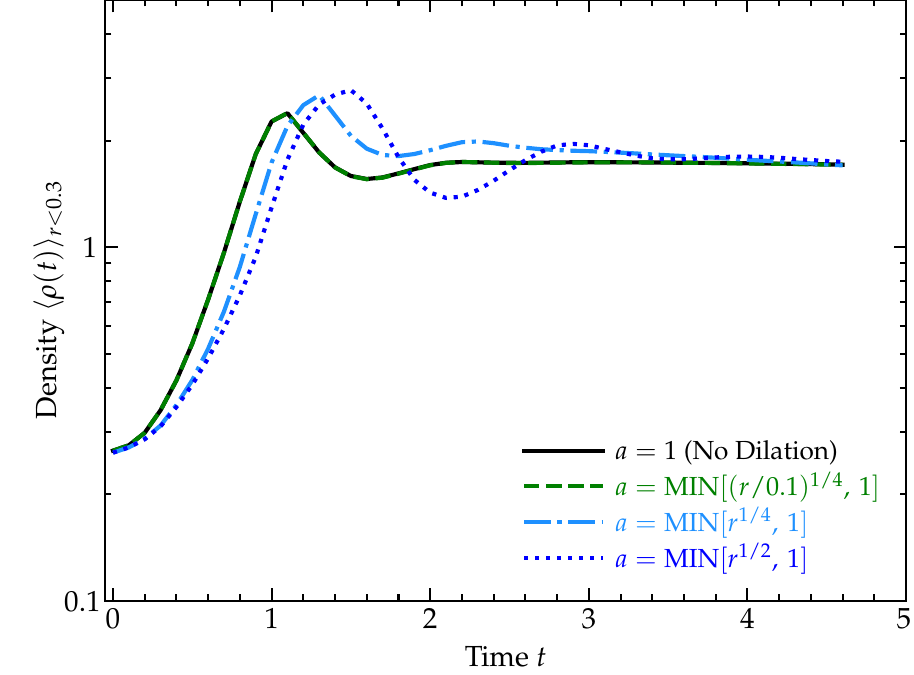} 
	\vspace{-0.2cm}
	\caption{3D non-equilibrium collapse/shock/hydrostatic equilibrium (Evrard) test (\S~\ref{sec:demo.test:evrard}). We initialize a spherical self-gravitating cold adiabatic cloud which collapses, shocks, bounces back, and oscillates before slowly reaching hydrostatic equilibrium. 
	\textit{Top:} Early time where all radii are near-maximally non-steady-state. A strong shock has formed but incompletely propagated through. The reference solution has $a=1$. 
	Dilation methods produce weakly different evolution at small radii where $a<1$. 
	However, these simply reflect the ``slowed'' evolution of the system, reflecting the exact solution at slightly earlier times (shown). The time lag can be smaller than $a$ (e.g.\ just $\sim 20\%$ in time here at $r \sim 10^{-3}$, where $a \sim 0.03$), owing to its dependence on $r$. 
	We compare a run which forces $\tilde{\bf S}\rightarrow 0$ in Eq.~\ref{eqn:hyper}, i.e.\ places $a$ inside $\nabla \cdot {\bf F}$ instead of with $D_{t} a$: this restores the traditional definitions of conserved quantities but produces qualitatively incorrect evolution.
	\textit{Middle:} Later time where the system is relaxing to equilibrium. The dilated and un-dilated methods relax to the same state.
	\textit{Bottom:} Time evolution of mean density inside $r<0.3$. Runs dilated on smaller scales agree with exact solutions, other runs are slowed down as expected before reaching equilibrium.
	\label{fig:evrard}}
\end{figure} 

\subsubsection{Non-Steady-State with Self-Gravity: Spherical Collapse}
\label{sec:demo.test:evrard}

Next we consider a significantly more challenging test problem involving strong non-equilibrium behavior and self-gravity, the \citet{evrard:1988.gas.collapse.problem} collapse problem \citep[widely used for testing codes with hydrodynamics plus self-gravity;][]{hernquistkatz:treesph,dave:1997.ptreesph,springel:gadget.public.release.paper,wadsley:2004.GASOLINE,springel:arepo,hopkins:lagrangian.pressure.sph,hopkins:gizmo}. In an open domain we initialize a self-gravitating sphere at low resolution ($30^{3}$ equal-mass elements) with mass $M=1$, radius $R=1$, polytropic index $\gamma=5/3$, initial density $\rho(r)=M/(2\pi R^{2} r)$ for $r<=R$ and $\rho=0$ for $r>R$, and thermal energy per unit mass $u=0.05$ (much less than gravitational energy). We integrate using the default second-order MFM and tree-gravity solver in {\small GIZMO}, as in \citet{hopkins:gizmo}. In this problem, the gas initially free-falls to $r\rightarrow 0$ under self-gravity, but the temperature increases and a strong shock forms in the center and the inner infall regions undergo a ``bounce'' and launch a shock back out through the infalling outer sphere. Eventually the shocks propagate throughout the system and it reaches a virial equilibrium. We consider several choices for $a$: $a=1$ (no dilation); $a={\rm MIN}[(r/0.1)^{1/4},\,1]$; $a={\rm MIN}[(r/0.1)^{1/2},\,1]$; $a={\rm MIN}[r^{1/4},\,1]$; $a={\rm MIN}[r^{1/2},\,1]$; $a=1/(1+r^{-1/2})$ -- many of these produce nearly-identical results, so we just show a subset for illustration. 

Traditionally the problem is compared at $t=0.8$, after a strong shock forms but well before the system virializes, so it is highly out-of-equilibrium (in fact, it is ``maximally'' non-equilibrium in the sense that the time-derivative $D_{t} {\bf U}$ terms are of the same magnitude as all other terms, $\nabla \cdot {\bf F}$, ${\bf S}$, in the evolution equations). We do so in Fig.~\ref{fig:evrard}. Recall, the methods here \textit{do not} ensure correct non-steady-state, non-equilibrium behavior. In fact the non-equilibrium behavior should be different, because we have fundamentally modified the evolution equations $D_{t} {\bf U}$ by multiplying by a radius-dependent $a({\bf x},\,t)$. So unsurprisingly, there are deviations from the exact solution on scales where $a \ll 1$ (i.e.\ in a radius-dependent fashion). What is important, however, is that these deviations appear largely \textit{as expected}. Specifically, the dilation factor $a(r) \ll 1$ at small $r$ should slow down evolution at those small radii, meaning they take longer to reach equilibrium (as is well-known and well-studied for e.g.\ RSL methods, see discussion in \citealt{skinner:2013.athena.reduced.c.implementation}). Looking at the time evolution of $\langle \rho(<r) \rangle$, we see that indeed the leading-order effect is that smaller $a$ produces more significant slowdown/lag, as it should. We see that if we consider slightly later times, the inner regions where the dilation factor was applied have indeed ``caught up'' to the expected solution. And also at radii where $a\rightarrow 1$, even in runs where $a<1$ at smaller radii, the solution traces the expected behavior (i.e.\ the effects are effectively local to where $a < 1$). So the algorithm is not systematically biasing the evolution in this test, merely slowing down the local high-rate-of-change variations on the way to local steady-state, precisely as it is designed to do in order to enable larger timesteps. 

At second order, we do see a small but expected effect where the extrema of the non-equilibrium $\rho$ as it relaxes are slightly enhanced for stronger scalings of $a(r)$. These result from the implicit $\nabla \ln{a}$ terms that appear when the system has not reached local steady-state. Consider: at early times here, the outer envelope (at radii where $a\rightarrow 1$) is collapsing at its ``full'' speed, but the ``bounce'' propagating back out is slowed by $a(r)<1$, so by the time it reaches the radii shown (the maximum in $\rho(r<0.3,\,t)$), more mass has piled up. This is explicitly an effect of local relative differences in $a$, i.e.\ $\nabla \ln{a}$, when far from steady-state. One can immediately verify it is independent of the absolute magnitude of $a$ by simply multiplying all $a$ in the simulation by a constant, which is identical to rescaling the time axis.

We also compare at a later time $t=4.6$, in Fig.~\ref{fig:evrard}. By this time, at most radii the system has approached equilibrium, with a relaxed, steep $\rho \propto r^{-3}$ density profile. We see that our time-dilated approaches recover this equilibrium/steady-state solution. This is not trivial -- while it is obvious that the methods here should \textit{preserve} the correct steady-state solution, it is not obvious that one can always ensure \textit{reaching} said steady-state solution from a non-equilibrium state which differs dramatically and approaches that state in a highly non-monotonic manner (as in this problem). Moreover, in many codes, the steady-state of this problem is considered quite challenging to preserve, because it relies on hydrostatic equilibrium between pressure forces and self-gravity, calculated in operator-split manner via different solvers. For specific Lagrangian methods like MFM this is not generally a problem \citep[see][]{hopkins:gizmo}, but it is not trivial to ensure the same is true with time dilation -- we need to ensure the two forces are indeed dilated appropriately, and the no effective source terms appear in the evolution equations, in order to ensure that we achieve the correct steady-state solution. This is demonstrated if we artificially set $\tilde{\bf S}\rightarrow 0$: per \S~\ref{sec:conservative}, this leads to incorrect dynamical and steady-state solutions. Note also the extreme dynamic range involved: with just $N=30^{3}$ resolution elements we represent a factor of $> 10^{4}$ in radius and $>10^{8}$ in density.

\begin{figure}
	\centering
	\includegraphics[width=0.98\columnwidth]{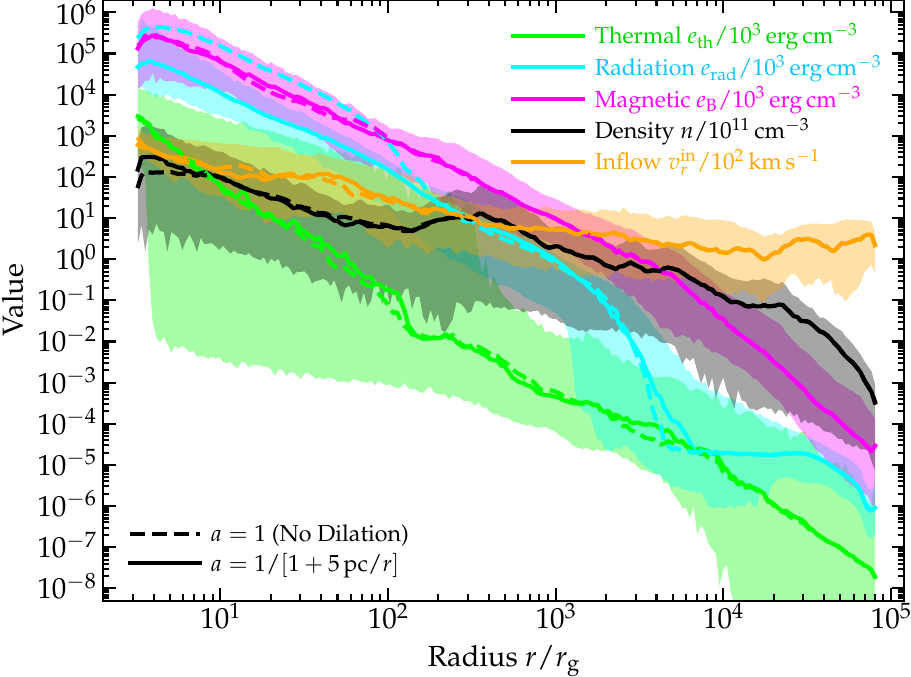} 
	\includegraphics[width=0.98\columnwidth]{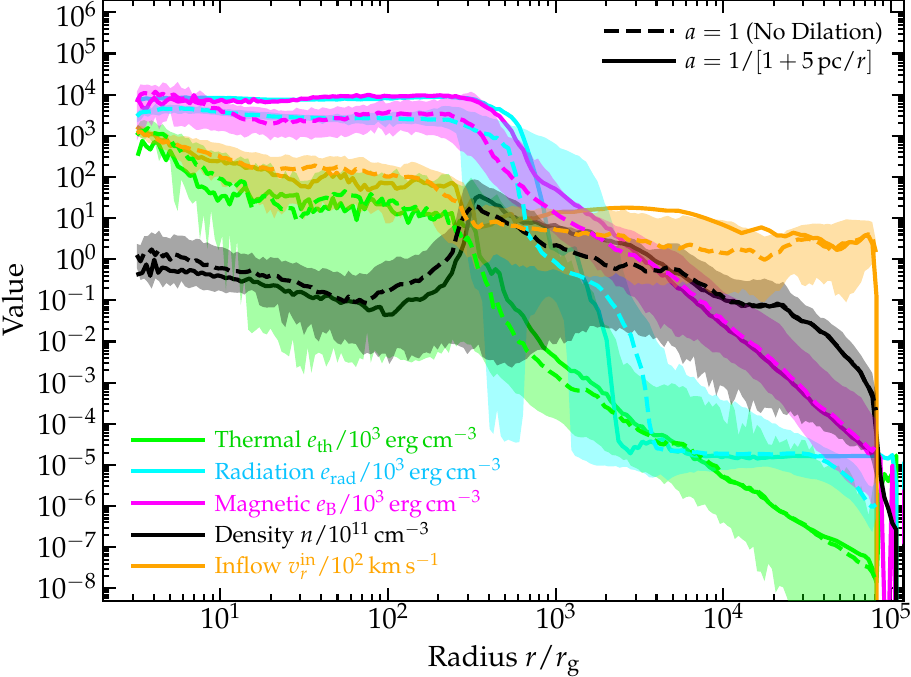} 
	\includegraphics[width=0.94\columnwidth]{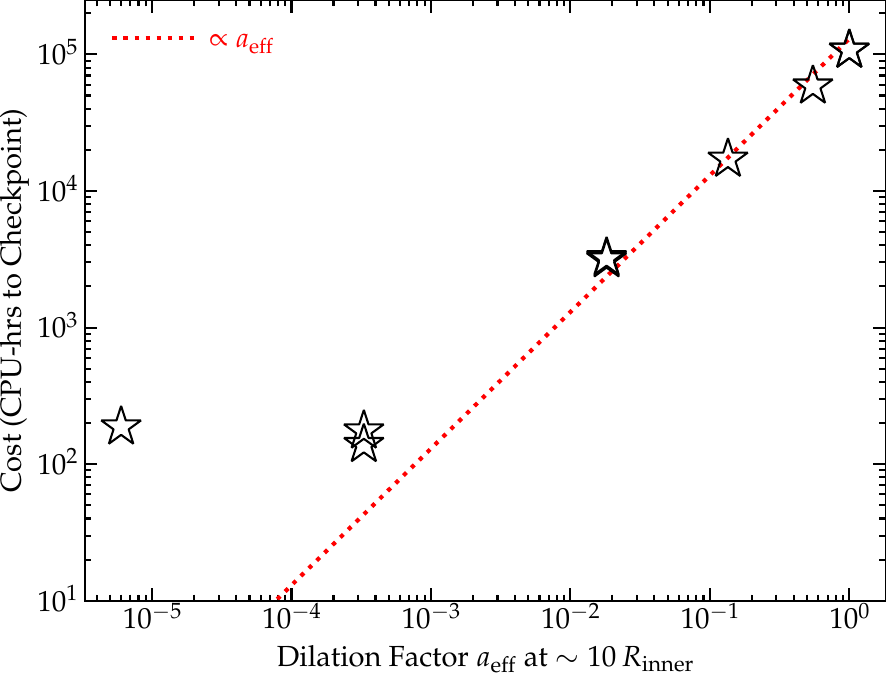} 
	\caption{Re-runs for a duration $\sim 10^{4}\,G M_{\rm BH}/c^{3}$ of a multi-physics, multi-scale simulation of quasar accretion (\S~\ref{sec:demo:agn}).
	\textit{Top:} Radial profiles (in $r_{g} \equiv 2\, G M_{\rm BH}/c^{2}$) of different quantities ($90\%$ range \textit{shaded}; median \textit{lines}), in the original simulation (\textit{dashed}) and a re-run with time dilation (\textit{solid}). The time is chosen so there is a steady-state disk down to the ISCO.
	\textit{Middle:} Same but chosen at a different time when the system goes into a strong radiation-pressure driven outburst with a MAD-like inner cavity.
	\textit{Bottom:} CPU cost to completion of the runs tested with different $a(r)$, in terms of an ``effective'' $a_{\rm eff}$ at $\sim 10\,$times the inner boundary (see text for details). The time-dilated simulations appear to behave similarly to their un-dilated counterparts, but are orders-of-magnitude less expensive. 
	\label{fig:agn}}
\end{figure} 

\subsubsection{Non-Equilibrium Magnetic Jet Launching: Collapsing Core}
\label{sec:demo.test:collapse_core}

Finally, we test a strongly non-equilibrium, magnetically dominated problem where Poynting flux is dynamically important and numerical divergence control is known to be critical. We adopt the collapsing-core MHD protostellar jet problem from \citet{hopkins:gizmo} (their Section 3.11 and Figs.~29-35), in which a rotating, self-gravitating core collapses to form a disk that winds up magnetic field lines and launches a non-relativistic jet. The flow is globally far from steady state, though locally portions of the jet and inner disk can be quasi-steady.

\begin{figure}
	\centering
	\includegraphics[width=\columnwidth]{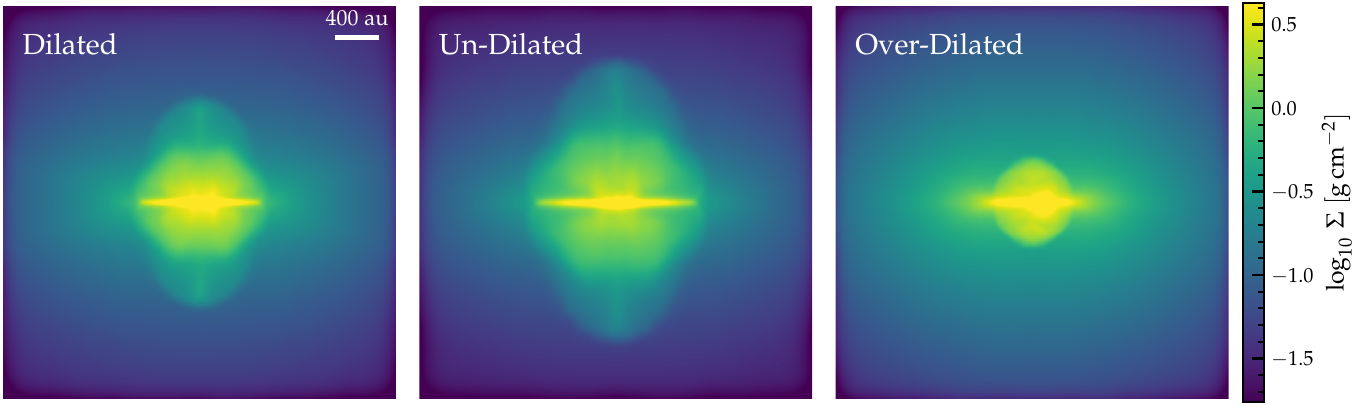}
	\caption{Projected density at $t=1.5\,t_{\rm ff}$ for the collapsing-core MHD jet test of \citet{hopkins:gizmo} (\S~\ref{sec:demo.test:collapse_core}). The three panels compare: no dilation (\textit{center}); a moderate dilation run with $a(r)$ ramped on linearly beginning at $0.9\,t_{\rm ff}$ over $\Delta t = 0.5\,t_{\rm ff}$ to $a(r)= \sqrt{r/r_{\rm max}}$ interior to $r_{\rm max}=50\,$au (\textit{left}); and an over-dilated case where the same profile is applied earlier, starting at $0.7\,t_{\rm ff}$ and reaching full strength by $0.9\,t_{\rm ff}$ (\textit{right}). The moderate dilation run reproduces the key jet-launching morphology with a somewhat less advanced jet (owing to the `slow-down' of its launch), while the over-dilated run is significantly delayed and shows a visually obvious asymmetry in the jet-launching region, consistent with the known sensitivity of this problem in divergence-cleaning schemes (like GIZMO used here) to tiny $\nabla\!\cdot\!{\bf B}$ errors (which arise because we ``stretch'' the central timesteps to exceed outer timesteps too-early, before the central core collapses).}
	\label{fig:collapse_core}
\end{figure}

We compare three runs at $t = 1.5\,t_{\rm ff}$: a reference case with no dilation; a moderate dilation case in which the dilation factor is ramped on linearly beginning at $0.9\,t_{\rm ff}$ and reaches $a(r)= \sqrt{r/r_{\rm max}}$ interior to $r_{\rm max}=50\,$au over $\Delta t = 0.5\,t_{\rm ff}$; and an over-dilated case where the same profile begins at $0.7\,t_{\rm ff}$ and reaches full strength by $0.9\,t_{\rm ff}$. The corresponding projected density comparison is shown in Fig.~\ref{fig:collapse_core}.

The moderate dilation run looks similar to the undilated reference, but is less evolved: the jet has not propagated as far by this global time, as expected when the central regions are effectively slowed. The over-dilated run is significantly delayed and develops a noticeable asymmetry in the jet-launching region. This occurs because the system had not yet reached sufficiently high central densities when the aggressive dilation was applied, so the inner core took longer timesteps than the outer regions. This makes it effectively impossible for the \citet{dedner:2002.divb.cleaning.scheme} divergence-cleaning scheme in GIZMO to effectively transport and damp $\nabla \cdot {\bf B}$ errors. But this problem is well-known to be extremely sensitive to such errors (part of its utility for divergence-cleaning schemes): even tiny $\nabla \cdot {\bf B}$ errors are rapidly amplified by the collapse, and these produce (via the \citealt{powell:1999.8wave.cleaning} control term needed for formal stability) violations of linear momentum conservation which lead to a symmetry-breaking displacement of the core. This test therefore demonstrates both that a carefully-chosen dilation scheme can preserve the essential jet morphology and Poynting-flux dominated behavior, while overly aggressive dilation can degrade the solution in a numerically sensitive MHD flow.

\subsection{Multi-Scale, Multi-Physics AGN}
\label{sec:demo:agn}

We now consider an example in a real, multi-physics, multi-scale problem. Specifically we consider the ``FORGE'd in FIRE'' simulations from \citet{hopkins:superzoom.analytic,hopkins:superzoom.disk,hopkins:superzoom.overview,hopkins:superzoom.imf,hopkins:superzoom.agn.disks.to.isco.with.gizmo.rad.thermochemical.properties.nlte.multiphase.resolution.studies,kaaz:2024.hamr.forged.fire.zoom.to.grmhd.magnetized.disks}. 
These are initially cosmological simulations which zoom in, refining continuously, to a major accretion episode onto a supermassive black hole in a high-redshift galaxy (at cosmological redshift $z\sim 4.5$), following the gas flows from Mpc to $\sim 10\,G M_{\rm BH}/c^{2}$ scales. They evolve a wide range of physics \citep[from][]{hopkins:2013.fire,hopkins:fire2.methods,hopkins:fire3.methods,grudic:starforge.methods} including: magneto-hydrodynamics; non-equilibrium multi-phase ionized/atomic/molecular/dust thermo-chemistry and cooling; multi-group/band radiation-hydrodynamics with opacities coupled to the thermochemical evolution; self-gravity; star formation with explicitly star-by-star and stellar-population based stellar evolution; stellar feedback in the form of protostellar jets, winds, radiation, and supernovae; dark matter; and more. Our goal here is not to delve into the science results or details of the simulations, so we do not discuss these further -- rather, we wish to use these as an extreme validation case for the methods here, similar in spirit to the supermassive black hole accretion problems considered in \citet{cho:2024.multizone.grmhd.sims.bondi.flow.lowaccrate} and \citet{guo:2025.grmhd.cyclic.zoom.modeling}. 

Specifically, in Fig.~\ref{fig:agn}, we compare several re-simulations, where we restart one of the simulations in \citet{hopkins:superzoom.agn.disks.to.isco.with.gizmo.rad.thermochemical.properties.nlte.multiphase.resolution.studies} from the same snapshot, and evolve forward with otherwise identical ``full'' physics and numerical methods as in that paper, varying only the choice of $a$. The reference case is $a=1$ again. We compare a few variations of $a$, most following 
$a^{-1} = 1 + (r_{0}/r)^{\zeta}$ with $r_{0}=5\,$pc (the BH radius of influence), and $\zeta = (1/6,\,1/3,\,\,2/3,\,1)$. We briefly tested other variants, with similar $a$ but imposing somewhat different $r_{0}$, or $a = {\rm MIN}[(r/r_{0})^{\zeta},\,1]$, or imposing a minimum/cutoff $a_{\rm min}$ at some small $r$, or using a cyclic $a$ as in \S~\ref{sec:self.valid}, and we tested restarting from three different points in time in the original simulation, but for the simple test here these all give similar results to the fiducial examples. 

First, for the duration that we are able to run the un-accelerated simulation ($a=1$), we compare the evolution of the central $\dot{M}$ as a measure of the local quasi-steady state and reasonable behavior. Other diagnostics from \citet{hopkins:superzoom.agn.disks.to.isco.with.gizmo.rad.thermochemical.properties.nlte.multiphase.resolution.studies}, e.g.\ the density/temperature/ionization/velocity profiles, give similar agreement, indicating that this is a reasonable application. Note the difference in the radiation energy density at small $r$ owes to the timing of a flare which occurs in both simulations, but is delayed to slightly after the time shown in the dilated case.

Second, we compare cost-to-completion for the (relatively short) test runs here. As expected, this scales strongly with the number of timesteps needed. Given the finite duration and resolution/inner boundary of the simulations we target, there is a point where reducing the minimum $a$ no longer decreases the simulation runtime: overheads, communication, and the collective cost of cells in larger timebins eventually dominate the total, and given the physical constraints we enforce on the timesteps and $a$, the timesteps can only be increased to a certain point. Where this will saturate is obviously problem-dependent. 
In the full fidelity simulation, CPU costs (given the extremely short timesteps near-horizon) limit the physical duration of the original simulation in \citet{hopkins:superzoom.agn.disks.to.isco.with.gizmo.rad.thermochemical.properties.nlte.multiphase.resolution.studies} to $\sim 10\,$days or $\sim 10^{4}\,G M_{\rm BH}/c^{3}$ (run over months of wall-clock time). But in one week of simulation wall-clock time (beyond the test shown above) we were able to advance the system to $\sim 5$\,years or $2 \times 10^{6}\,G M_{\rm BH}/c^{3}$, a speedup of a factor of $\sim 5000 = 0.5 \times 10^{4}$. In that run, the minimum dilation factor adopted was $a_{\rm min} \sim 10^{-4}$, so the ideal speedup if there were no overheads and all the CPU work was associated strictly with the smallest timestep would be a factor $1/a_{\rm min} = 10^{4}$. This is therefore only a factor of a couple below the ``ideal'' speedup (somewhat better than the scaling in Fig.~\ref{fig:agn}, owing to the larger box, higher resolution, and longer run-time of this experiment), consistent with the notion that indeed, the vast majority of the compute time was spent on smallest timesteps. Note that by reducing the depth of the timestep hierarchy (therefore the number of timesteps where only a tiny fraction of cells are active), this hugely improves the load-balancing and efficiency of the simulations (beyond just reducing total number of timesteps to completion). Over the duration of the extended time-dilated run, the system exhibits an approximately constant time-averaged mass flux $\langle \dot{M} \rangle$ as a function of radius (fluctuating about the expected accretion rate), consistent with the accretion flow having reached a statistical steady-state at each scale. The time-averaged radial energy flux profile likewise converges, with the energy carried by the jets/winds at large radii balancing accretion power at small radii to within the expected variability. Detailed analysis of the long-duration run physics will be presented in a forthcoming study; the purpose here is to demonstrate the method's practical viability for such applications.

\section{Discussion: Tradeoffs of these Approaches}
\label{sec:tradeoffs}

We now briefly summarize some advantages and disadvantages (compared to the dilation methods proposed here) of different methods commonly used in the literature to deal with similar problems. 

\subsection{Practical Guidelines for Choosing $a({\bf x},\,t)$}
\label{sec:tradeoffs:choosing.a}

We briefly offer some practical guidance for selecting $a$, beyond the formal criteria in \S~\ref{sec:criteria}. In general, for problems dominated by a single central potential (e.g.\ accretion onto a point mass), a natural starting point is $a(r) = 1/(1 + (r_{0}/r)^{\zeta})$ or $a(r) = {\rm MIN}[(r/r_{0})^{\zeta},\,1]$ with $r_{0}$ of order the transition radius where the timestep begins to become prohibitively small, and $\zeta \sim 1/3$--$1$ controlling the aggressiveness of the dilation. Smaller $\zeta$ gives a gentler scaling (less speedup but safer for systems further from steady-state), while $\zeta \sim 1$ gives near-maximal speedup consistent with the ordering requirement in \S~\ref{sec:criteria} for Keplerian systems. In our experience, the tradeoff between fidelity and speedup is relatively forgiving: as long as the criteria in \S~\ref{sec:criteria} are satisfied and the system is reasonably close to local steady-state, the results are insensitive to the precise functional form of $a$ at the factor-of-two level (see the Bondi and AGN tests in \S~\ref{sec:demo}). When in doubt, a conservative approach is to start with a gentle $\zeta$ and progressively increase it, comparing diagnostics (e.g.\ time-averaged fluxes) against a short brute-force reference run. For problems with multiple special regions, the dilation factor can be defined as $a({\bf x}) = \min_{i} a(|{\bf x}-{\bf x}_{i}|)$ over the set of special points ${\bf x}_{i}$, each with their own profile. The key practical ``sweet spot'' is typically where the minimum $a$ is small enough that the timestep hierarchy is substantially compressed (bringing the minimum and maximum timesteps closer together), but not so small that the de-dilation criteria in \S~\ref{sec:self.valid} are frequently triggered.

\subsection{Versus ``Brute Force'' Methods}
\label{sec:tradeoffs:force}

Ideally, one would of course simply ``do it all'' -- simulating all scales at once with high resolution and the correct timesteps, integrated over the global evolution time. This has many obvious advantages over approximate methods like those proposed here. But the obvious disadvantage (motivating our methods in the first place) is computational cost: it is simply not possible to simulate the entire range of timescales -- or even close to it -- in many astrophysical contexts (see \S~\ref{sec:intro}). For example, state-of-the-art ``full fidelity'' simulations are, at present, often limited to something like $\sim 10^{7}$ dynamic range between the shortest and longest evolved timescales (e.g.\ of order $\sim 10^{7}-10^{8}$ timesteps; see e.g.\ \citealt{daa:20.hyperrefinement.bh.growth,guszejnov:environment.feedback.starforge.imf,applebaum:2021.highres.justice.league.highres.mw.mass.galaxy,hopkins:superzoom.overview,hopkins:superzoom.agn.disks.to.isco.with.gizmo.rad.thermochemical.properties.nlte.multiphase.resolution.studies,cho:2024.multizone.grmhd.sims.bondi.flow.lowaccrate}). While some impressive exceptions exist, these are still many orders of magnitude away from numerous problems where the salient dynamic range in time approaches $\sim 10^{15}-10^{17}$. But we stress that \textit{the methods here are in no way a replacement for full-fidelity simulations}. Recall, there is no absolute guarantee of convergence of our methods to correct solutions in truly general problems (where local steady-state may not even exist). The methods here absolutely require calibration, validation, and convergence testing for different types of problems (just as done with the methods that inspire them, like RSL methods). For these tests, full-fidelity simulations are required. Even if these can only be run for a fraction of the dynamical time at a given scale, they provide key physical insights and confidence that methods like those here can be trusted for certain predictions under certain conditions.

\subsection{Versus ``Simulation Stitching (Subgrid)'' Methods}
\label{sec:tradeoffs:stitching}

The more traditional approach to multi-scale simulations -- including most of our own work -- has been simulation ``stitching,'' as reviewed in \S~\ref{sec:intro}. 
Note that this category can include explicit simulation-to-simulation maps, machine-learning/artificial intelligence (ML) driven models trained on simulations or observations, or more traditional ``sub-grid'' models from either fitting simulations/observations or highly simplified analytic models of some scales. 
Advantages of stitching include (1) it is easy to reach arbitrarily high resolution at low CPU cost, in sub-simulations arbitrarily small volumes and times; (2) it is (often) comparatively ``easy'' to set up; (3) all scales can be simulated on their true timesteps (just not in the same simulation); (4) it enables more precise control of initial/boundary conditions and physics assumptions; (5) one can ignore, at each scale, physics unimportant for those scales; and (6) it is possible to use different codes/numerical methods independently optimized to each scale/physics. 

But there are a number of drawbacks to stitching methods which include: (1) the scale (and physics) separation assumed is often artificial or nonexistant; (2) small and large scales are strongly coupled so one cannot be sure the solutions are globally valid; (3) one can ``miss'' critical parameter space/IC/BCs owing to reliance on prior assumptions regarding inner/outer scales; (4) one still cannot run the small-scale problem for anything like the dynamic range of time of large scales; (5) some highly nonlinear dynamics (e.g.\ global modes, accreting or blowing out all the mass in the sub-domain being simulated, etc.) typically cannot be captured; (6) numerical errors often arise when stitching between different codes; (7) the stitching almost always entails substantial loss of information from simulations of one domain applied to another, as only certain variables/maps/information are applied by subgrid/ML models; and (8) the formal parameter space explored by the space of stitched simulations is vast and extremely multi-dimensional, so stitching almost always involves extrapolation into un-explored space (e.g.\ the out-of-distribution problem) rather than interpolation. 

All of these are reasons to think of the methods here as akin to a ``better sub-grid model'' or more accurately an improved method for more self-consistent simulation stitching. The methods here are not doing anything worse (in terms of accuracy), from a numerical or physical point of view, compared to a traditional stitching or subgrid model approach, and they capture numerous effects which would otherwise be missed. But given the advantages of stitching noted above, methods like those here will not and should not replace stitching in all contexts. Even where it is possible to apply the methods here (e.g.\ to the test problem of BH accretion above), it remains infeasible to refine on e.g.\ every BH in a large cosmological volume at all times. They can, however, be used to inform traditional subgrid/stiching methods: to ask e.g.\ what the key parameter space is, where the models can be trusted, where the artificially-imposed scale-separation is (or is not) an acceptable approximation, etc.

\subsection{Versus ``Iterative/Cyclic Refinement'' Methods}
\label{sec:tradeoffs:cyclic}

As noted above, iterative/cyclic zoom techniques are really just a special case of the methods here, with $a({\bf x},\,t)$ being a series of step functions between $0$ and $1$ in both space and time, rather than continuous functions. So there is no real disadvantage of the methods here, by comparison. However, by generalizing those techniques, we introduce a number of advantages for more general applications including (1) removing artificial ``breaks'' in spatial scale, allowing for problems where there is no discrete scale-separation; (2) allowing for a smooth/continuous change of timestep; (3) ensuring differences in $a$ between neighboring cells are small, greatly reducing errors in the local dynamics and conservation; (4) enabling much greater flexibility for different desired physics to be scaled differently, treating different domains and arbitrary refinement schemes, etc; (5) making the method much easier to implement in arbitrary numerical schemes, especially quasi-Lagrangian or ALE or moving-mesh methods; (6) removing the need for refinement/de-refinement of the inner grid (i.e.\ transient ``coarse-graining'' of the grid), so low-order errors and loss of information associated with these operations are removed (the resolution of the refined sub-domains can remain high throughout the simulation); and (7) enabling flexible application to implicit/explicit time-integration methods.

\subsection{Versus ``Projective Integration'' and HMM Methods}
\label{sec:tradeoffs:projective}

Projective integration methods, as an example of the ``equation free'' methods for multiscale time integration, and some closely-related heterogeneous multiscale methods (HMM) essentially take some fixed number $N_{0}$ of ``regular'' timesteps $\Delta t$ for fast-timescale physics, then compute the mean rate of change for variables over that set of timesteps $\langle \partial_{t} {\bf U}_{i} \rangle_{N_{0}}$, and use this to drift the variables or project the solution over some much longer timescale $\Delta t_{\rm proj} \gg \Delta t$, before repeating \citep[see e.g.][and references therein]{tretiak:2022.multiscale.methods.twotimescale.average.and.project}. The method proposed here is attempting to accomplish something similar in spirit, and could be cast in the framework of such models by considering it to take alternating full/projection steps ($N_{0}=1$) with the assumption that the time-averages down by a factor $a$ in the projection step. But really the approaches (as designed by default) have different applications. Most projective integration and other closely related highly multiscale methods like those discussed in \citet{kevrekidis:2009.equation.free.approaches.to.highly.multiscale.simulation.problems.review,weinan:2011.principles.multiscale.problems.simulations.hmm.projective.amr.more} work well in problems where the rapid-timescale variation averages down to a mean change which is indeed very small, i.e.\ turbulence in an isotropic medium, where $\langle {\bf v} \rangle \rightarrow \mathbf{0}$ over sufficiently-long timescales and $\| \langle \partial_{t} {\bf U}_{i} \rangle_{N_{0}} \|  \ll \| \partial_{t} {\bf U}_{i} \| $, justifying the long projection timestep. But this is not the same as the statistical steady-state we invoke here. For example, if there is a steady mass flux into one part of a sub-domain with small timesteps (e.g.\ a disk) and steady (offsetting) mass flux out (e.g.\ a bipolar wind/jet), then the system can obey statistical steady state but have a large mass flux rate into/out of any single cell, which would give unphysical results (e.g.\ negative densities) if extrapolated over a very large timestep with most projective integration methods. Or more simply, in a Lagrangian code, a steady-state circular disk involves a large nominal Lagrangian derivative in the position ${\bf x}$ or velocity ${\bf v}$ of a cell, which cannot be treated with standard projective integration methods. So there are many problems for which those methods will simply not work. However, for problems where the mean rate-of-change of all the evolved ${\bf U}_{i}$ is slow (in the ``fast'' physics domain/resolution), then those methods may well be preferred to ours, as they can give more faithful higher-order time evolution and can be applied to volume-filling physics.

Note that many other multiscale methods also exist in the literature. Some, like those in \citet{tang:2024.small.timestep.forced.to.equilibrium.sph}, assume a strict scale-separation and activate short timesteps occasionally but \textit{force} the system to an exact local equilibrium state by adding artificial numerical damping terms, which is useful if the equilibrium should be exact (and some of its characteristics are known), but is not applicable to general, slowly-evolving and non-uniquely scale-separated problems.

\subsection{Versus Implicit/Semi-Implicit Methods and Super-Timestepping}
\label{sec:tradeoffs:implicit}

For some problems -- for example, purely-local (within a single cell, no neighbor information used) calculations of thermochemistry -- implicit or semi-implicit methods allow for vastly larger timesteps (compared to pure explicit methods). Likewise certain problems such as simple diffusion operators admit ``super-timestepping'' schemes to allow somewhat larger stable explicit timesteps. Of course this should still be used where optimal, and our methods here are no substitute for such approaches. But a further gain for some problems can be realized by applying both -- the dilation terms are agnostic to how the time update is done and can be applied to implicit or explicit methods equally. However, implicit (or super-timestepped) methods cannot be reliably applied with arbitrarily large timesteps to all equations/problems, especially those of greatest interest here. For example, even if one could feasibly write the equations $D_{t}{\bf U} = \mathcal{F}({\bf U})$ in some form that ensured formal numerical stability for arbitrarily large timesteps (e.g.\ with some implicit Euler-like ${\bf U}^{(n+1)} = {\bf U}^{(n)} + \mathcal{F}^{(n+1)}({\bf U}^{(n+1)},\,..., t^{(n+1)})\,\Delta t$), this (a) would almost certainly prove intractable (requiring all updates be global simultaneous solves over all variables ${\bf U}_{i}$ for all resolution elements), and (b) would not be \textit{accurate} (despite being stable) for any timestep larger than the local dynamical evolution time (which was the rate-limiter of interest in the first place for our methods).

\section{Summary}
\label{sec:summary}

We have presented an extremely simple and flexible approach for time-advancement of highly multiscale problems, which is itself a generalization of approaches already developed for radiation/neutrino/cosmic ray dynamics, divergence-cleaning, cyclic/iterative zoom/refinement approaches, hard-binary collisional N-body dynamics, and similar in spirit to projective integration and other ``equation free multiscale methods'' in the literature. This amounts to applying a variable (but continuous) time-dilation factor $a({\bf x},\,t)$ to the dynamics as a function of space and time, or equivalently to stretching each physical timestep by $1/a$ along the global integer timeline. 
Like all of those other methods, this relies fundamentally on the assumption that the subdomains where $a < 1$ can be treated as in statistical steady-state or quasi-equilibrium on timescales of order their global timeline step (i.e.\ evolve secularly, in an ensemble sense). Indeed it recovers correct solutions in steady-state ($D_{t} \rightarrow 0$) by construction. When those conditions are approximately satisfied, we show that this allows for orders-of-magnitude speedup in advancing extremely multi-scale problems to large global times. However, we stress that these are no replacement for traditional ``full fidelity'' ($a=1$ everywhere) simulations: the approach we propose approach is changing the dynamical equations -- it is fundamentally a ``hack'' -- and so must be validated against traditional simulations on different problems where possible.

We show how conservative and nonconservative forms of the method can be applied; derive criteria $a$ must obey to ensure good numerical and physical behavior; describe associated ``activation'' criteria to avoid pathological mismatches of timesteps between neighbor cells and additional dilated-timestep criteria that must be obeyed; and outline variations of the method for cases where there is a clear ``fast-slow dynamics'' separation for some equations in time, or in space, or in different subsets of physics/variables being evolved. Given the dependence of the method validity on statistical steady-state behavior in subdomains, we present methods to quasi-periodically ``de-dilate'' to reset the true dynamics either at scheduled points in the simulation or adaptively based on on-the-fly criteria that check for ``sufficiently non-steady-state'' dynamics. This allows for some adaptive ``self-correction'' of the simulations using these methods. 

We highlight test problems (both new here and examples in the literature which are equivalent to special cases of the methods here), and applications to modern multi-physics, multi-scale problem of accretion flows onto and feedback (winds, radiation) flowing out from a supermassive black hole. Despite steady-state only being approximate we show the method appears consistent with full-dynamics simulations, and enables an enormous effective speedup. We show examples demonstrating effective increases in CPU efficiency of $\sim 10^{3}-10^{6}$, enabling correspondingly longer maximum timescales to be reached. 
This therefore seems like a promising method to study systems which are intrinsically extremely multi-scale in time as well as space, especially in cases where there are well-defined special regions or sub-domains of the global spacetime simulation domain in which the extreme dynamic range is needed.
We discuss the relation of these methods to other methods in the literature for treating such extreme dynamic range problems, and highlight advantages or disadvantages of each.

\begin{acknowledgements}
We thank Haiyang Wang, Minghao Guo, and Doron Kushnir for helpful conversations. Support for PFH was provided by a Simons Investigator Grant. ERM acknowledges support by the National Science Foundation under grants NSF-AST2508940 and NSF-PHY2309210. Numerical calculations were run on NSF TACC allocation AST21010.
\end{acknowledgements}

\bibliographystyle{mn2e}
\bibliography{ms_extracted}

\begin{appendix}

\section{Connection to Physical Time Dilation and Interpretation of Source Terms}
\label{sec:gr}

In our method, the ``dilation factor'' $a$ is purely a numerical abstraction, and does not represent physical time dilation as would arise in general relativity (GR). However it is instructive to consider the GR connection, as it helps elucidate the form of the equations we propose and the interpretation of source terms that arise in the strictly-conservative form of the equations.

Consider the GR(RM)HD equations, for (radiation-magneto)hydrodynamics in GR, in the 3+1 formulation that allows us to write them in terms of spacelike hypersurfaces described by the coordinates ${\bf x}^{i} = \{ 1,\,2,\,3\}$ each of which is at a constant value of a timelike coordinate ${\bf x}^{0} = \{ t \}$ {with unit normal $n_\mu = (-\alpha,0,0,0)$, where $\alpha$ is the lapse (i.e., time dilation) function. } Importantly, $\alpha$ represents a coordinate transformation to a set of observers, so can in principle be an arbitrary function.
{The four-dimensional metric can then be expressed via the line element}
$d s^{2} = -(\alpha^{2} - \beta_{i} \beta^{i})\,dt^{2} + 2\,\beta_{i} dx^{i} dt + \gamma_{ij} dx^{i} dx^{j}$,
{where $\gamma_{ij}$ is the induced metric on the spatial hypersurface, }
$\gamma \equiv {\rm det}(\gamma_{ij})$ and $\beta$ is the ``shift'' vector. 
Then the equations can be written in the form \citep{banyuls:1997.gr.3plus1.equations,font:2008.grmhd.equations.methods.review}: 
\begin{align}
\label{eqn:gr} \frac{\partial \sqrt{\gamma} {\bf \mathcal{U}}({\bf w}) }{\partial t}
+ \frac{\partial \sqrt{\gamma} \alpha {\bf \mathcal{F}}^{i}({\bf w}) }{\partial {\bf x}^{i}} = \sqrt{\gamma} \alpha {\bf \mathcal{S}({\bf w})}
\end{align} 
in terms of conservative variables ${\bf \mathcal{U}}$, fluxes ${\bf \mathcal{F}}$, sources ${\bf \mathcal{S}}$, which are functions of some primitive variables ${\bf w}$.

Taking the limit of Minkowski space, Eq.~\ref{eqn:gr} reduces to the special-relativistic SR(RM)HD equations, or considering weak gravity and slow speeds, to the Newtonian (RM)HD equations with external gravity. 

Instead consider the limit of a flat metric and slow speeds, but retain the term $\alpha$ in generality to represent time dilation. It is straightforward to show that the standard GRMHD source terms in ${\bf \mathcal{S}({\bf w})}$ are metric curvature terms (even if some components of them are sometimes written explicitly with $\alpha$ in them), and this can also be seen from the traditional Newtonian expansion (where e.g.\ the momentum source term becomes the $\nabla \Phi$ gravitational acceleration), so we can drop them but for generality we retain some ${\bf S}_{\rm flat}$ to represent Newtonian or numerical source terms (e.g.\ injection of photons on a grid). In terms of the Newtonian variables ${\bf U}$ and fluxes ${\bf F}$ we have: 
\begin{align}
\label{eqn:gr.to.newton} \frac{\partial {\bf U} }{\partial t} + \nabla \cdot \left( \alpha {\bf F}  \right) = \alpha {\bf S}_{\rm flat}
\end{align} 
or, after some simple algebra:
\begin{align}
\label{eqn:gr.to.newton.alt} \frac{1}{\alpha} \frac{\partial {\bf U} }{\partial t} + \frac{1}{\alpha}\nabla \cdot \left( \alpha {\bf F} \right)  = {\bf S}_{\rm flat} \ , \\
\nonumber \frac{1}{\alpha} \frac{\partial {\bf U} }{\partial t} + \nabla \cdot {\bf F}  = {\bf S}_{\rm flat} - {\bf F} \cdot \nabla \ln{\alpha}\ .
\end{align} 
This is identical to our proposed dilation factor, except for the source terms ${\bf F} \cdot \nabla \ln{\alpha}$. {We can understand this term by considering the weak-field limit of a gravitating point mass where $\alpha = \sqrt{1+2\Phi/c^2}$ so that $\rho c^2\nabla \ln{\alpha} \simeq \rho \nabla \Phi$. In other words, since general relativity couples time dilation to the presence of a gravitating mass, any time dilation must entail a gravitational force.}

These make it such that steady-state solutions are no longer independent of $\alpha$, which suggests they have a physical meaning: indeed they are associated with gravitational red/blue-shifts. 

To see this more clearly, consider the advection equation: $\alpha^{-1} \partial_{t} \rho + \nabla\cdot(\rho{\bf u}) = -\rho{\bf u}\cdot\nabla\ln{\alpha}$. In a divergence-free flow ($\nabla \cdot (\rho{\bf u})$ small, so the Lagrangian change in the volume of a test parcel is negligible), this becomes $\partial_{t} \ln{\rho} = -\alpha {\bf u}\cdot \nabla \ln {\alpha}$ or, in terms of the comoving derivative with a parcel along ${\bf u}$ (recalling $d{\bf x}/dt = \alpha\,{\bf u}$ here), $d\ln{\rho} = -d\ln{\alpha}$ or $\rho \propto \alpha^{-1}$. This makes it obvious that the source terms in $\nabla \ln{\alpha}$ represent gravitational blueshift/redshift: as a parcel of fixed volume falls into a region of small $\alpha$ (strong dilation), its density and therefore (since it has fixed volume) total rest mass-energy density go up $\propto \alpha^{-1}$, as appropriate for a gravitational blueshift (exactly how $\alpha$ appears in the metric). And vice-versa upon outflow. This is not the desired behavior in a strictly Newtonian system. 

So in our formulation, these source terms should be dropped as well (along with the curvature terms), keeping only the ``trivial'' coordinate term. Taking that limit and replacing $\alpha \rightarrow a$ (to make it clear this is not a \textit{physical} time dilation), we have: 
\begin{align}
\label{eqn:gr.to.us.final} \frac{1}{a} \frac{\partial {\bf U} }{\partial t} + \nabla \cdot {\bf F}  = {\bf S}_{\rm flat}  \ , 
\end{align} 
or (if we wish to rewrite the equations in a manifestly antisymmetric form): 
\begin{align}
\label{eqn:antisymm} \frac{1}{a} \frac{\partial {\bf U} }{\partial t} + \frac{\nabla \cdot \left( a  {\bf F} \right)}{a}   = {\bf S}_{\rm flat} + {\bf F} \cdot \nabla \ln{a} \ .
\end{align} 
Note Eq.~\ref{eqn:gr.to.us.final} and Eq.~\ref{eqn:antisymm} are identical. Both make it plain that so long as $a$ does not depend explicitly on ${\bf w}$ in local interactions, the hyperbolic character of the original equations is preserved. There are simply different use cases where one or the other is more numerically advantageous. Eq.~\ref{eqn:antisymm} makes it clear, by comparison to Eq.~\ref{eqn:gr.to.newton.alt}, that the ``source term'' in $\nabla \ln{a}$ which appears when we put $a$ \textit{inside} the divergence $\nabla \cdot (a {\bf F})$ serves to cancel the otherwise spurious blueshift/redshift terms that would appear in the evolution of conserved quantities.

\section{Lagrangian vs.\ Eulerian Derivatives}
\label{sec:lagrangian}

Briefly, in Lagrangian codes, it is important to note that the salient Lagrangian derivative is $d_{t} \equiv \partial_{t} + {\bf v}_{{\rm mesh},\,i} \cdot \nabla$, where ${\bf v}_{{\rm mesh},\,i} \equiv d{\bf x}_{{\rm mesh},\,i}/dt$ is defined by the rate-of-change of the mesh-generating points defined \textit{on the simulation coordinate grid} $({\bf x},\,t)$. In quasi-Lagrangian methods like SPH, MFM, and some variants of moving mesh and MFV methods, this is set to the ``fluid'' velocity, but here it really means to the same definition of time derivative of the center-of-mass of a Lagrangian fluid element in lab coordinates, which is now ${\bf v}_{{\rm mesh},\,i} \rightarrow a \langle {\bf u}\rangle = a {\bf u}_{i}$. Thus the Lagrangian derivative of interest becomes $d_{t} \rightarrow \partial_{t} + a {\bf u} \cdot \nabla$, and we have 
\begin{align}
\frac{1}{a}\frac{\partial {\bf U}}{\partial t} \rightarrow \frac{1}{a} \frac{d {\bf U}}{ d t} - \left( {\bf u} \cdot \nabla \right) {\bf U}
\end{align}
which means that we have the ``usual'' comoving equations outside of the $a^{-1} \partial_{t} {\bf U}$ or $a^{-1} d_{t} {\bf U}$ (written in terms of ${\bf u}$ without explicit $a$ dependence). 
The same follows for e.g.\ the conservative comoving derivative operator $\hat{D}_{t}{\bf U} \equiv \rho\, d_{t} [{\bf U}/\rho]$, 
\begin{align}
\frac{1}{a} \frac{ \partial {\bf U}} {\partial t} \rightarrow \frac{1}{a} \frac{\hat{D}{\bf U}}{\partial t} - \nabla \cdot \left( {\bf u} {\bf U} \right)\ .
\end{align}
 In any case, this justifies our statement in \S~\ref{sec:method} that we can apply the dilation factor (as long as we do so \textit{systematically}) to either $\partial_{t}{\bf U}$ (as is generally useful in Eulerian methods) or to $d_{t}{\bf U}$ or $\hat{D}_{t} {\bf U}$ (as are commonly used in Lagrangian methods) -- in either case the remaining equation is identical and the $a^{-1}$ appears only in front of the salient time derivative.

\section{Enforcing Divergence Constraints}
\label{sec:divergence}

Numerically enforcing a divergence constraint of the form $\nabla \cdot {\bf q} = 0$ is often important, in particular for incompressible flows (${\bf q}={\bf u}$; see examples in \citealt{gresho:1990.vortex,tiwari:2003.finite.pointset.method,liu:2005.finite.particle.method,basic:2022.lagrangian.methods.dealing.with.sharp.edges.novel.method.sharp.boundaries.for.sph.mfm.to.use}) or MHD (${\bf q}={\bf B}$; \citealt{evans:1988.constrained.transport,powell:1999.8wave.cleaning,balsara:2001.amr.mhd.review.failures.of.powell.and.other.subtraction.methods,dedner:2002.divb.cleaning.scheme,mocz:2014.constrained.transport.mhd,hopkins:cg.mhd.gizmo}). It is not obvious that our method preserves these constraints. For example, consider the Eulerian continuity and induction equations. In the incompressible case, $\partial_{t}\rho = - \nabla \cdot (\rho {\bf u}) = -\rho \nabla \cdot {\bf u} = 0$ is trivially maintained if we replace $\partial_{t} \rho \rightarrow a^{-1}\partial_{t} \rho$ (i.e.\ the same constraint $\nabla \cdot {\bf u} = 0$ applies). However consider the Eulerian induction equation for ${\bf B}$, $\partial_{t} {\bf B} |_{a=1} = -\nabla \times {\bf E}$, which trivially (analytically) ensures $\partial_{t} \nabla \cdot {\bf B} = 0$. If we modify this per Eq.~\ref{eqn:method}, we have $\partial_{t} {\bf B} = -a\,\nabla \times {\bf E}$ so $\partial_{t} \nabla \cdot {\bf B} = \partial_{t} {\bf B} |_{a=1} \cdot \nabla a$, which vanishes when the system is in steady-state but is not guaranteed to vanish otherwise. Numerically, even in the $a=1$ case, it is well-known these constraints require special care as they can be violated numerically, in an unstable manner, even when the equations preserve them. And one must be careful to define the precise numerical definition of the divergence operator which relates to the term producing said instability (one can always find a different numerical divergence operator which vanishes, without actually removing the instabilities or errors produced by the method). 

We note that the treatment of ${\bf B}$ here (applying the dilation factor uniformly to all evolution equations, including the induction equation) differs from the procedure adopted in \citet{cho:2024.multizone.grmhd.sims.bondi.flow.lowaccrate} and \citet{guo:2025.grmhd.cyclic.zoom.modeling}, where the magnetic field at zone boundaries requires special treatment owing to their discontinuous activation/deactivation of zones. In our continuous formulation, $a$ multiplies the induction equation $\partial_{t} {\bf B} = -a\,\nabla \times {\bf E}$ uniformly, so the magnetic field evolution is slowed in precisely the same manner as all other variables, and no special boundary conditions for ${\bf B}$ are needed at interfaces. This is a direct advantage of the continuous approach.

However in popular methods for treating these errors, our approach is easily incorporated. There are two qualitative approaches to the problem. First, vector potential or constrained-transport or related methods do not directly evolve cell-centered values of ${\bf q}$, but rather define ${\bf q}$ as a \textit{derived} quantity from some other field (e.g.\ an effective vector potential ${\bf A}$, where ${\bf B} = \nabla \times {\bf A}$ and $\partial_{t} {\bf A} = -{\bf E}$). The dilation factor should therefore be applied to the vector potential/electric field step, which preserves the divergence free-character of ${\bf q}$ to the same accuracy as the original method. This is no more difficult to code than any other constrained-transport or vector-potential method which already allows for differing timesteps in neighboring cells.

In our MHD test problems (\S~\ref{sec:demo:agn}), we have verified that the normalized magnetic divergence $h\,|\nabla \cdot {\bf B}|/|{\bf B}|$ (where $h$ is the local kernel/cell size) remains at levels $\lesssim 10^{-2}$--$10^{-3}$ throughout the dilated simulations, comparable to the levels in the un-dilated ($a=1$) reference runs. This is consistent with our expectation from the analysis above that the additional divergence error scales as $|\Delta x\,\nabla \ln{a}| \ll 1$ when the smoothness criteria on $a$ are satisfied. We have not attempted a vector potential formulation for these tests, as {\small GIZMO} primarily employs divergence-cleaning methods \citep{hopkins:cg.mhd.gizmo}; however, as noted above, the vector potential approach should be straightforwardly compatible with the dilation method (applying $a$ to the evolution of ${\bf A}$ rather than ${\bf B}$ directly), and would by construction maintain $\nabla \cdot {\bf B} = 0$ to machine precision. The main disadvantage of vector potential methods --- that gauge freedom complicates the evolution and derivative operations can reduce numerical accuracy --- is unrelated to the dilation and would persist regardless of $a$.

The primary alternative to these are some form of divergence-cleaning or projection methods which attempt to damp or project out such errors after each step. 
In strictly-Lagrangian methods, for example, it is common to explicitly work with the forms $\hat{D}_{t} \rho = 0$ (i.e.\ to not explicitly integrate $\rho$ but calculate it implicitly from the updated cell configurations ${\bf x}_{i}$, where $\dot{\bf x}_{i} = a\,{\bf u}_{i}$), and for MHD $\hat{D}_{t} {\bf B} = ({\bf B} \cdot \nabla)\,{\bf u}$ (for which the cell-centered-and-averaged $\langle {\bf B}_{i} \rangle = {\bf B}_{i}$ will not manifestly respect $\hat{D}_{t} \nabla \cdot {\bf B} = 0$). So both already produce $\nabla \cdot {\bf q} \ne 0$, and the update with e.g.\ $\dot{x} = a\,{\bf u}$ or $\hat{D}_{t} {\bf B} = a\,({\bf B} \cdot \nabla) {\bf u}$ can be calculated just like any other update, with the cleaning/projection applied immediately following, to ensure the erroneous terms are removed. To ensure cleaning can operate effectively, we should compare the maximum amplitude of the divergence error: $| \Delta t D_{t} \nabla {\bf B} |  / |{\bf B} / \Delta x | \sim | \nabla a \Delta x | \sim a | \Delta x \nabla \ln{a} | \ll 1$, which is satisfied so long as the basic constraints on $a$ from \S~\ref{sec:criteria} are met.

\end{appendix}

\,
\\
\,
\\

\end{document}